\documentclass[floats,floatfix,amssymb,prd,twocolumn,superscriptaddress,nofootinbib]{revtex4-1}

\usepackage[english]{babel}
\usepackage[utf8]{inputenc}
\usepackage{amsmath}
\usepackage{mathbbol}
\usepackage{amssymb}
\usepackage{graphicx,amsfonts}
\usepackage{float}
\usepackage{epsfig}
\usepackage[normalem]{ulem}
\usepackage[colorlinks=true,
linkcolor=blue,
urlcolor=blue,
citecolor=blue]{hyperref}
\usepackage{bm}
\usepackage{mathrsfs}
\usepackage{mathtools}
\usepackage{enumerate}
\usepackage{amsthm}
\usepackage{bbm}
\usepackage{comment}
\usepackage{physics}
\usepackage{url}
\usepackage{upgreek}
\usepackage[svgnames]{xcolor}
\usepackage[title]{appendix}

\begin{document}

\title{Excitation factors for horizonless compact objects:\\ long-lived modes, echoes, and greybody factors}

\author{Romeo Felice Rosato}
\affiliation{Dipartimento di Fisica, Sapienza Università di Roma \& INFN, Sezione di Roma, Piazzale Aldo Moro 5, 00185, Roma, Italy}

\author{Shauvik Biswas}
\affiliation{Indian Institute of Technology, Gandhinagar, Gujarat 382055, India}

\author{Sumanta Chakraborty}
\affiliation{School of Physical Sciences, Indian Association for the Cultivation of Science, Kolkata 700032, India}

\author{Paolo Pani}
\affiliation{Dipartimento di Fisica, Sapienza Università di Roma \& INFN, Sezione di Roma, Piazzale Aldo Moro 5, 00185, Roma, Italy}


\begin{abstract}
We present an analytical and numerical investigation of the quasinormal excitation factors of ultracompact horizonless objects. These systems possess long-lived quasinormal modes with extremely small imaginary parts, originating from the effective cavity between the photon sphere and the object’s interior. We show that the excitation of such modes is strongly suppressed, scaling with the imaginary part of their frequency, and therefore they contribute to the waveform only at very late times. This hierarchy naturally explains the structure of echo signals: the prompt ringdown is dominated by standard light-ring modes, the early echoes arise from moderately damped cavity modes, and only the latest echoes are governed by long-lived modes.
Building on this, we propose a practical ringdown waveform model that combines ordinary black-hole quasinormal modes with cavity modes, capturing the complexity of the ringdown of horizonless ultracompact objects.
We further demonstrate that the combination of small excitation factors and weak damping enhances the robustness of long-lived modes against localized perturbations, in contrast to the spectral instabilities affecting standard black-hole quasinormal modes. Finally, we extend the analysis of greybody factors to exotic compact objects and wormholes, showing that they remain stable under small deformations of the effective potential and thus represent robust observables. Our results provide a unified framework for understanding excitation, stability, and echoes in ultracompact horizonless objects, with direct implications for their spectral properties and gravitational-wave signatures.
\end{abstract}

\maketitle 

\tableofcontents

\section{Introduction}
Black-hole (BH) perturbation theory~\cite{Regge:1957td,Chandrasekhar:1985kt} is recently undergoing a phase of rapid advancement (see~\cite{Berti:2025hly,Carullo:2025oms} for recent reviews). This revival has been largely driven by the first gravitational-wave~(GW) detections of remnant ringdowns~\cite{LIGOScientific:2016aoc,LIGOScientific:2020iuh,LIGOScientific:2025epi,LIGOScientific:2025obp} and by the prospect of BH spectroscopy~\cite{Dreyer:2003bv,Detweiler:1980gk,Berti:2005ys,Gossan:2011ha} as a probe of gravity in the strong-field, dynamical regime~\cite{LIGOScientific:2021sio,Berti:2015itd,Berti:2018vdi,Cardoso:2019rvt}.

The next decade promises a qualitative leap in precision, with current detectors reaching design sensitivity and with the advent of next-generation interferometers such as LISA~\cite{LISA:2024hlh}, Einstein Telescope~\cite{ET:2019dnz,Kalogera:2021bya,Hild:2010id,Branchesi:2023mws,Abac:2025saz}, and Cosmic Explorer~\cite{LIGOScientific:2016wof,Essick:2017wyl,Evans:2023euw}. These facilities are expected to measure ringdowns at the sub-percent level~\cite{Berti:2016lat,Bhagwat:2021kwv,Bhagwat:2023jwv}, enabling unprecedented tests of General Relativity and of the nature of compact objects~\cite{Cardoso:2019rvt}.

A central goal of BH spectroscopy is the extraction of multiple quasinormal modes (QNMs)~\cite{Vishveshwara:1970zz,Kokkotas:1999bd,Berti:2009kk,Konoplya:2011qq} from the post-merger remnant as it approaches stationarity. 
In General Relativity, the entire QNM spectrum of a BH is uniquely determined by its mass and spin, allowing for stringent null-hypothesis tests of the theory~\cite{Isi:2019aib,Franchini:2023eda}, of the remnant’s nature~\cite{Maggio:2020jml,Maggio:2021ans,Maggio:2023fwy}, and of its surrounding environment~\cite{Barausse:2014tra,Cardoso:2021wlq,Cardoso:2022whc,Destounis:2022obl}. The push toward accurate ringdown modeling has, however, uncovered additional layers of complexity, including environmental imprints~\cite{Barausse:2014tra,Barausse:2014pra,Cheung:2021bol,Berti:2022xfj,Biswas:2023ofz,Singha:2023lum}, modified boundary conditions or near-horizon structures~\cite{Cardoso:2016rao,Cardoso:2016oxy,Cardoso:2017cqb,Abedi:2020ujo}, spectral instabilities~\cite{Nollert:1996rf,Daghigh:2020jyk,Jaramillo:2020tuu,Destounis:2021lum,Gasperin:2021kfv,Boyanov:2022ark,Jaramillo:2022kuv,Sarkar:2023rhp,Destounis:2023nmb,Arean:2023ejh,Cownden:2023dam,Destounis:2023ruj,Courty:2023rxk,Boyanov:2023qqf,Cao:2024oud,Cardoso:2024mrw,Ianniccari:2024ysv,Cai:2025irl}, nonlinear effects~\cite{Gleiser:1995gx,Gleiser:1998rw,Ioka:2007ak,Nakano:2007cj,Brizuela:2009qd,Pazos:2010xf,Ripley:2020xby,Loutrel:2020wbw,Sberna:2021eui,Cheung:2022rbm,Mitman:2022qdl,Kehagias:2023ctr,Perrone:2023jzq,Cheung:2023vki,Redondo-Yuste:2023ipg,Redondo-Yuste:2023seq,Yi:2024elj,Zhu:2024dyl,Zhu:2024rej}, the role of overtones~\cite{Giesler:2019uxc,Bhagwat:2019dtm,Baibhav:2023clw}, and late-time tails~\cite{Price:1972pw,Gundlach:1993tp,Barack:1998bw,DeAmicis:2024not,DeAmicis:2024eoy,Rosato:2025rtr}.

In parallel with QNM-based analyses, an alternative line of investigation has emerged, focusing on BH greybody factors (GFs)~\cite{Oshita:2022pkc,Oshita:2023cjz,Okabayashi:2024qbz}—real, frequency-dependent functions describing the tunneling probability through the BH effective potential~\cite{Hawking:1975vcx}. GFs have been shown to capture the spectral amplitude of the ringdown at frequencies above the fundamental mode~\cite{Oshita:2022pkc,Oshita:2023cjz,Rosato:2024arw,Okabayashi:2024qbz}, to remain robust under small deformations of the system, and to make contact with quantities that can be expressed as QNM superpositions, even in the presence of spectral instabilities~\cite{Rosato:2024arw,Oshita:2024fzf}.

In this work we focus on the ringdown of ultracompact, horizonless objects, often referred to as BH mimickers or exotic compact objects (ECOs)~\cite{Cardoso:2019rvt,Maggio:2021ans,Maggio:2020jml,Bambi:2025wjx}. It has been shown that, if the remnant is sufficiently compact~\cite{Cardoso:2016rao,Cardoso:2016oxy,Cardoso:2017cqb}, the prompt ringdown signal following the merger is universal, since it is governed by the local structure of the effective potential of null geodesics. Putative near-horizon structures, which can be modeled in terms of a reflectivity coefficient at the object’s surface~\cite{Mark:2017dnq,Maggio:2017ivp}, show up only at late times in the form of repeated and modulated GW echoes~\cite{Cardoso:2016rao,Cardoso:2016oxy,Abedi:2016hgu,Cardoso:2017cqb,Abedi:2016hgu}. These echoes involved long-lived QNMs and for a fully relativistic and observationally relevant example of a binary-merger of ultra-compact horizonless objects, see Ref.~\cite{Siemonsen:2024snb}.The time delay between echoes depends on the compactness, while the relative amplitude of successive echoes encodes the reflectivity. Recently, it was further demonstrated that signals from ECOs also develop a late-time tail coinciding with that of BHs~\cite{Rosato:2025rtr}. From a broader perspective, the long-lived modes discussed here are closely analogous to cavity modes in leaking optical resonators~\cite{Lalanne:2018, Sheikh:2022cud}, providing a natural language for studying resonant scattering in open systems.

Several models of ECOs have been investigated over the years, including wormholes~\cite{Visser:1995cc}, constant-reflectivity ECOs~\cite{Maggio:2017ivp,Mark:2017dnq,Maggio:2020jml}, and objects with Boltzmann reflectivity~\cite{Oshita:2019sat}.
Here we provide an analytical and numerical characterization of the QNMs excitation  for ECOs. Quasinormal excitation factors (QNEFs)~\cite{Leaver:1986gd,Berti:2006wq,Silva:2024ffz} encode a universal, initial-data–independent measure of the relative excitation of QNMs, arising from the poles of the Green’s function that propagates small perturbations of the geometry. QNEFs do not depend on the perturbation details; when combined with knowledge of the initial data they yield the quasinormal excitation coefficients, which quantify the actual QNM content of a waveform given a specific source.

To the best of our knowledge, this work presents the first attempt toward the computation of QNEFs for ECOs and BH mimickers. The QNM spectrum of such objects is characterized by the presence of long-lived modes, which arise because a horizonless object has a non-vanishing reflectivity, thereby creating an effective cavity between its interior and the gravitational effective potential barrier. We show that QNM excitation is strongly correlated with the small imaginary part of these mode: \emph{the smaller the damping, the weaker the excitation}. This observation has significant implications for the physics of ECOs. In particular, it clarifies how cavity modes generate the sequence of echoes in the waveform. Long-lived modes have extremely small imaginary parts (i.e., nearly real frequencies), resulting in very weak excitation and contributing to the signal only at very late times, once all other modes have exponentially decayed. The cavity spectrum, however, is broader: it also contains modes with larger imaginary part than the long-lived branch, though still smaller than those of the Schwarzschild QNM spectrum. These modes are responsible for generating the first echoes observed after the prompt ringdown. This is discussed in detail in Sec.~\ref{excitationfactors}, while in Sec.~\ref{sec:echoreconstruction} we show how to reconstruct the echo signal from a superposition of QNMs with the corresponding  excitation factors.

We also analyze the stability of the mode spectrum, which is intrinsically connected to the excitation factors. Indeed, an important feature of  QNMs is their extreme sensitivity to small perturbations of the system~\cite{Nollert:1996rf,Barausse:2014tra,Daghigh:2020jyk,Jaramillo:2020tuu}, be them in the background geometry or in the boundary conditions. This suggests that the QNM spectrum of a BH could be drastically altered by environmental effects~\cite{Barausse:2014tra,Barausse:2014pra,Cheung:2021bol,Berti:2022xfj} or near-horizon structures~\cite{Cardoso:2016rao,Cardoso:2016oxy,Cardoso:2017cqb,Abedi:2020ujo}, although the prompt ringdown in the time domain is known to be far less affected~\cite{Cardoso:2016rao,Cardoso:2016oxy,Cardoso:2017cqb,Mirbabayi:2018mdm,Berti:2022xfj,Kyutoku:2022gbr}. We investigate how wormholes and ECOs respond to such deformations~\cite{Boyanov:2022ark,Destounis:2025dck}. As discussed in Sec.~\ref{stability}, we find that long-lived modes are remarkably stable under small perturbations of the effective potential, providing an analytical understanding that is in agreement with some recent numerical results~\cite{Destounis:2025dck}. As we shall show, this robustness originates from the fact that the small imaginary part of the mode, together with the corresponding suppression of its excitation factor, prevents any significant amplification of the system’s response. 

As a corollary, in Sec.~\ref{sec:stabilityGF} we further verify that the GFs of wormholes and ECOs—recently demonstrated to be stable against small deformations of the effective gravitational potential for BHs~\cite{Rosato:2024arw,Oshita:2024fzf}—remain stable under analogous perturbations for these objects as well. This point is particularly relevant, as GFs play a crucial role in modeling the frequency-domain spectral amplitude of GWs for both ECOs and wormholes, as recently shown~\cite{Rosato:2025byu}.\\

In the following, we use $G=c=1$ units and work exclusively in the four dimensional spacetimes. Further, we use a signature convention, such that the flat metric in Cartesian coordinates becomes $\eta_{\mu \nu}=\textrm{diag.}(-1,+1,+1,+1)$.
\section{Excitation factors of long-lived modes}\label{excitationfactors}
For ultracompact objects, the fundamental quasinormal QNMs possess a very small imaginary part, corresponding to long-lived oscillations that remain quasi-trapped between the object's interior and its photon sphere~\cite{Cardoso:2014sna}. 
In the case of wormholes, this behavior arises due to the cavity formed between the two photon spheres on the two sides of the wormhole throat.

In this section, we examine the excitation factors associated with these long-lived QNMs for ultracompact, non-rotating objects~\cite{Cardoso:2019rvt,Maggio:2021ans,Bambi:2025wjx}. We restrict to the non-rotating case and assume that the spacetime geometry outside the ultracompact object ---or at least beyond an effective radius $r_0$--- is well described by the Schwarzschild metric. In both cases, the system is governed by the one-dimensional radial wave equation:
\begin{equation}\label{Regge-Zerilli}
\Bigg[{d^2 \over dr_*^2} + \omega^2 - V_l(r)\Bigg] X_{lm\omega}=0\,,
\end{equation}
where $\omega$ is the frequency of the perturbation, the  tortoise coordinate $r_{*}$ is defined via $dr/dr_{*}=1-2M/r$, and $M$ is the mass of the object. The function $X_{lm\omega}$ represents the radial part of the perturbation, which may be scalar, electromagnetic, or gravitational in nature, and $V_l(r)$ denotes the corresponding effective potential. For gravitational perturbations, $V_l(r)$ corresponds to either the Regge-Wheeler or Zerilli potential, depending on the parity of the perturbation. 

The eigenvalue problem associated with the second-order differential equation~\eqref{Regge-Zerilli} requires two boundary conditions: one at spatial infinity and the other at the inner boundary of the object, which depends on the specific model under consideration. QNMs require an outgoing boundary condition at infinity:
\begin{equation}\label{boundary_refl/trasm}
 {X}_{lm\omega} \to   e^{+i\omega r_*}\,; \quad r_*\to+\infty\,,
\end{equation}
where a time dependence of the form $e^{-i\omega t}$ is assumed for the perturbations, so that the positive (negative) exponential corresponds to an outgoing (ingoing) wave. 

The inner boundary condition depends on the internal structure of the compact object and has to be discussed separately for each case. In general, it involves a superposition of outgoing and ingoing waves at the surface of the object, with the amplitude of the outgoing wave governed by the object's reflectivity.

In the following, we focus on two classes of ultracompact objects:
\begin{itemize}

    \item \emph{Schwarzschild-like wormholes}, modeled by joining two Regge-Wheeler\footnote{For concreteness, we focus on gravitational axial perturbations; however, the analysis applies more generally.} potentials at the origin. The effective potential in this case is given by~\cite{Visser:1995cc, Bueno_2018}
    \begin{equation}\label{wormholepot}
    V_l(r_*)=\theta(r_{*}-r_{*}^{0})W_{l}(r_{*})+\theta(r_{*}^{0}-r_{*})W_l(-r_{*})\,,
    \end{equation}
   with $W_l(r_*)$ being the Regge-Wheeler potential, and $r_*^0$ being the throat location in tortoise coordinates (while we shall denote the throat location in Schwarzschild coordinates as $r_{\rm throat}=r(r_*^0)$).

    \item \emph{Schwarzschild-like reflecting ECOs}, whose exterior geometry matches the Schwarzschild solution up to an effective radius $r_0$. For ultracompact objects, the compactness condition $r_0-2M \ll M$ holds. For external observers, the ECO is modeled by the \emph{surface reflection amplitude} $R_{\rm ECO}$ at its surface and its compactness $\epsilon=(r_{0}/2M)-1$. These two parameters capture all the information about the linear response of the ECO interior~\cite{Maggio:2017ivp, Mark:2017dnq, Maggio:2020jml}. Near this effective radius, the perturbing field behaves as~\cite{Mark:2017dnq, Maggio:2020jml, Chakraborty:2022zlq},
    \begin{equation}\label{ECObcs}
        \Psi\to e^{- {i} \omega (r_*-r_*^0)} + R_{\rm ECO}  e^{ {i} \omega (r_*-r_*^0)}\,, \quad r_*\to r_*^0
    \end{equation}
    where $r_*^0=r_*(r_0)$, the tortoise coordinate associated with the surface of the ECO.
\end{itemize}

The QNM problem can be regarded as a special case of the more general wave scattering problem, where the asymptotic behavior is
\begin{equation}\label{boundary_refl/trasm}
{X}_{lm\omega} \to A^{\rm in}_{lm\omega} e^{-i\omega r_*}+ A^{\rm out}_{lm\omega} e^{+i\omega r_*}\,; 
\quad 
r_*\to+\infty\,.
\end{equation}
In particular, QNM frequencies correspond to the complex roots, $\omega_n=\omega^R_n+i\omega^I_n$, of the equation
\begin{equation}\label{eqQNMS}
    A^{\rm in}_{lm\omega}\Big|_{\omega_n}=0\,.
\end{equation}

The boundary condition~\eqref{boundary_refl/trasm} also enables the definition of reflection and transmission amplitudes
\begin{equation}
    R_{lm}(\omega)={A^{\rm out}_{lm\omega} \over A^{\rm in}_{lm\omega}}\,,\qquad T_{lm}(\omega)={1 \over A^{\rm in}_{lm\omega}}\,.
\end{equation}
From the above quantities one can compute the QNEFs, namely~\cite{Berti:2006wq}
\begin{equation}\label{excitation}
    B_n={A^{\rm out}_{lm\omega} \over 2  \omega_n \gamma_{lmn}}\,,\quad \text{with}\quad \gamma_{lmn}={dA^{\rm in}_{lm\omega}\over d\omega}\Bigg|_{\omega_n}\,,
\end{equation}
where $n$ identifies the overtone number. The QNEFs should not be confused with the quasinormal excitation coefficients, which instead depend on the specific source $I_{lm\omega}(r_*)$ entering the inhomogeneous version of Eq.~\eqref{Regge-Zerilli}, and are defined as \cite{Berti:2006wq}
\begin{equation}\label{ex_coefficients}
C_n = B_n \int_{-\infty}^{+\infty} {\psi_n(r_*)\, I_{lm\omega}(r_*) dr_* \over A^{\rm out}_{lm\omega}}\,,
\end{equation}
where $\psi_n(r_*)$ denotes the solution of Eq.~\eqref{Regge-Zerilli} corresponding to the specific $n$-th QNM. While the excitation factors $B_n$ provide universal measures of QNM excitation, the coefficients $C_n$ quantify the excitation of a given mode by a given source.

In the following sections we compute the QNEFs $B_n$ both numerically and analytically. In the latter case the calculations are performed in the regime $|\omega|M \ll 1$, where Eq.~\eqref{Regge-Zerilli} can be solved analytically~\cite{Starobinskil:1974nkd}. Although restricted to this regime, these results provide useful insights into the general behavior of the QNEFs for the long-lived modes of wormholes and ECOs, while the full frequency range is addressed numerically.

We find it useful to introduce the \emph{cavity length}, $L$. 
In the case of an ECO, where a partially reflecting surface is located at $r_* = r_*^0$, this length corresponds to the distance between the surface at $r_*=r_*^{0}$ and the peak of the gravitational potential at $r_*=r_*^{\rm max}$, namely $L =|r_*^{0}-r_*^{\rm max}|\approx|r_*^{0}|$ (since for ultracompact objects the reflective surface lies at large negative tortoise coordinates compared to the maximum of the potential). 
In contrast, for a wormhole configuration, the cavity length corresponds to the distance between the two potential barriers.

\textbf{Wormhole case.}  
Using a transfer matrix approach~\cite{Ianniccari:2024ysv}, Ref.~\cite{Rosato:2025byu} demonstrates that, for wormholes, the ingoing wave amplitude $A^{\rm in}_{lm\omega}$ can be computed as 
\begin{equation}\label{eqAinwormhole}
    A^{\rm in}_{lm\omega}=\left(\alpha^{\rm in}_{lm \omega}\right)^2 e^{-i \omega L} - \left(\alpha^{\rm out,*}_{lm\omega}\right)^2 e^{i \omega L}\,,
\end{equation}
where $L$ represents the effective length of the cavity, whereas $\alpha^{\rm in}_{lm \omega}$ and $\alpha^{\rm out}_{lm \omega}$ correspond to the standard ingoing and outgoing wave amplitudes in the BH case\footnote{Namely, they correspond to the boundary conditions in Eq.~\eqref{boundary_refl/trasm} specialized to the BH case, where one simply replaces the symbol $A$ with $\alpha$ in that equation.}~\cite{Rosato:2025byu}. In this case, Eq.~\eqref{eqQNMS} takes the form
\begin{equation}
    e^{-2i \omega_n L} = \left(\frac{\alpha^{\rm out,*}_{lm\omega_n}}{\alpha^{\rm in}_{lm \omega_n}}\right)^2 = e^{2\log\left(\frac{\alpha^{\rm out,*}_{lm\omega_n}}{\alpha^{\rm in}_{lm \omega_n}}\right)} e^{-2\pi i n}\,, \quad n \in \mathbb{Z}\,.
\end{equation}
The  QNMs are therefore given by~\cite{Maggio:2018ivz}
\begin{equation}
    \omega_n=\omega^R_n+i\omega^I_n \sim { \pi n \over L}+\frac{i}{L} \log\left( \frac{\alpha^{\rm out}_{lm\omega_n}}{\alpha^{\rm in}_{lm \omega_n}} \right)\,,
\end{equation} 
with $n \in \mathbb{Z}$.
Note that long-lived modes appear in the low-frequency regime, in particular they are characterized by $|{\omega^I_n}| \ll |{\omega^R_n}| \ll M^{-1}$. In this limit, where the frequencies are approximately real, the ratio $\alpha^{\rm in}_{lm \omega} / \alpha^{\rm out}_{lm \omega}$ can be approximated as~\cite{Starobinskil:1974nkd}
\begin{equation}\label{strarobinky}
  \left|\frac{\alpha^{\rm out}_{lm\omega_n}}{\alpha^{\rm in}_{lm \omega_n}}\right| = 1 - 2 \beta_{ls} \left(2M\omega_n^R\right)^{2l+2} + \mathcal{O}\left(\left(M\omega_n^R\right)^{2l+4}\right)\,,
\end{equation}
where $\beta_{sl} = \left(\frac{(l-s)!(l+s)!}{(2l)!(2l+1)!!}\right)^2$ and $s=(0,\pm1,\pm2)$ for scalar, electromagnetic, and gravitational perturbations, respectively. This leads to the expression~\cite{Cardoso:2019rvt}
\begin{equation}\label{eq:wormholeim}
     \omega^I_n \sim -\frac{2 \beta_{ls}}{L} \left(2M\omega^R_n\right)^{2l+2}\,,
\end{equation}
which clearly satisfies our requirement $|\omega^I_n| \ll \omega^R_n$ for any $n$.

Now consider the following Taylor expansion for $A^{\rm in}_{lm\omega}$ around the QNM frequency $\omega_n$,
\begin{equation}
    A^{\rm in}_{lm\omega_n} = \left.\frac{d A^{\rm in}_{lm\omega}}{d\omega}\right|_{\omega_n} \left(\omega - \omega_n\right) + \mathcal{O}\left((\omega - \omega_n)^2\right)\,.
\end{equation}

Using Eq.~\eqref{eqQNMS} and the approximation~\eqref{strarobinky}, we find that, at $\omega = \omega_n$,
\begin{equation}
 \frac{d A^{\rm in}_{lm\omega}}{d\omega} \sim -2iL \left(\alpha^{\rm out,*}_{lm \omega}\right)^2 e^{i \omega_n L}\,,
\end{equation}
so that
\begin{equation}
 \left|\frac{d A^{\rm in}_{lm\omega}}{d\omega}\right| \sim 2L |\alpha^{\rm out}_{lm \omega}|^2 = \frac{L}{2 \beta_{ls}} \left(2M{\omega^R_n}\right)^{-2l-2}\,.
\end{equation}
This yields
\begin{equation}\label{eqproduct}
    \left|{\omega^I_n} \frac{d A^{\rm in}_{lm\omega}}{d\omega} \right|_{\omega_n} \sim 1 + \mathcal{O}\left((M{\omega^R_n})^2\right)\,.
\end{equation}
We also need to compute the outgoing wave amplitude $A^{\rm out}_{lm\omega}$. 
Ref.~\cite{Rosato:2025byu} shows that, for wormholes,
\begin{equation}\label{eqAoutwormhole}
    A^{\rm out}_{lm\omega}=\alpha^{\rm in}_{lm \omega} \alpha^{\rm out}_{lm \omega} e^{-i \omega L} - \alpha^{\rm in,*}_{lm \omega} \alpha^{\rm out,*}_{lm \omega} e^{i \omega L}\,.
\end{equation}
If one considers the squared absolute value of Eq.~\eqref{eqAoutwormhole} and implements Eq.~\eqref{eqQNMS} for the wormhole case, it follows that
\begin{equation}\label{eqAoutwormhole2}
    \left|A^{\rm out}_{lm\omega}\right|^2=\left(| \alpha^{\rm out}_{lm \omega}|^2  - |\alpha^{\rm in}_{lm \omega} |^2\right)^2\,,
\end{equation}
implying that $|A^{\rm out}_{lm\omega_n}|=1$ at the QNM frequency, since in the limit of nearly real frequencies it holds that $|\alpha^{\rm in}_{lm \omega}|^2 - |\alpha^{\rm out}_{lm \omega}|^2 = 1$~\cite{Rosato:2025byu}.
We can now compute the QNEFs  $B_n$ defined in Eq.~\eqref{excitation} to be 
\begin{equation}
    |B_n| \sim 
    {|{\omega^I_n}| \over 2 |\omega_n|} \sim 
   \beta_{ls} \frac{2M}{L}(2M\omega_n^R)^{2l+1}\,,
    \label{BnomegaI}
\end{equation}
showing that the excitation factors satisfy $|B_n| \ll 1$ for long-lived modes, since they scale with ${\omega^I_n}$.
We stress that this result is valid only when ${\omega^I_n} \ll{\omega^R_n} < M^{-1}$. Nevertheless, it provides an interesting insight into how the QNEFs scale with the imaginary part of the relative QNM.

Equation~\eqref{BnomegaI} indicates that long-lived modes are only weakly excited during the ringdown phase. However, owing to their very small imaginary parts, they experience little damping and may eventually dominate the signal once the more strongly excited but shorter-lived modes have decayed.
As a result, after the initial prompt ringdown, long-lived modes can re-emerge and produce so-called echoes (beatings of the cavity frequencies) with amplitudes much smaller than that of the initial signal.

\begin{figure*}
    \centering    \includegraphics[width=0.9\linewidth]{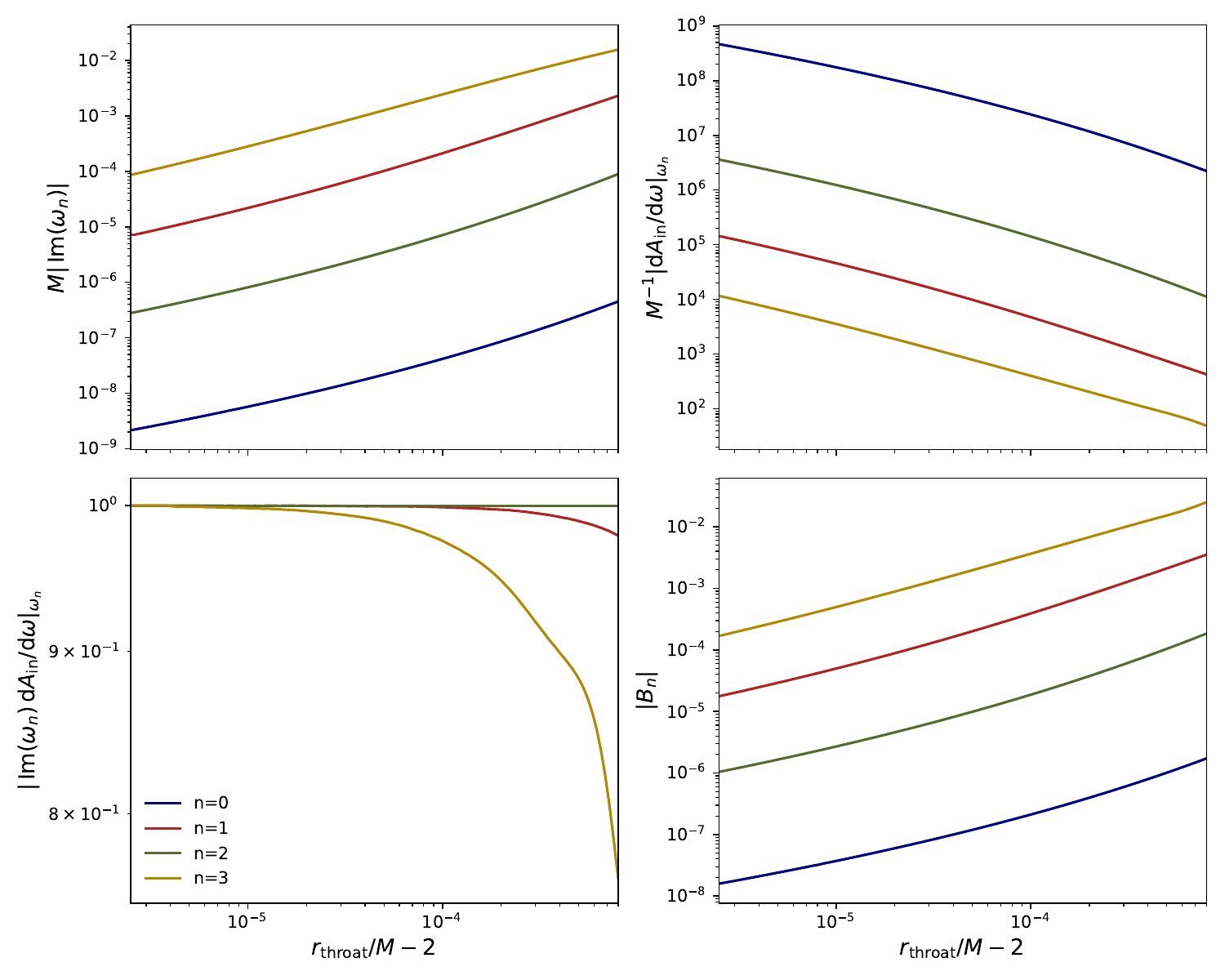}
   \caption{ 
   QNMs and excitation factors for axial gravitational perturbations with $l=2$ in the wormhole scenario.
 \emph{Top left panel:} imaginary part of the first four long-lived modes, as a function of the throat location (horizontal axis). \emph{Top right panel:} absolute value of $dA^{\rm in}/d\omega$ for the corresponding modes, using the same color coding. $dA^{\rm in}/d\omega$ becomes very large for long-lived modes, indicating that these modes are extremely suppressed (see Eq.~\eqref{excitation}). 
 \emph{Bottom left panel}: the product ${\omega^I_n}\, dA^{\rm in}/d\omega$, which approaches unity when the throat location satisfies $\frac{(r_{\rm throat} - 2M )}{2M}\ll 1$. This confirms the analytical predictions of Eq.~\eqref{eqproduct}. \emph{Bottom right panel}:  $|B_n|$ for the considered modes, using the same color coding. The QNEF is smaller for smaller imaginary part of the mode.
}
    \label{fig:excitationwormhole}
\end{figure*}

The results presented in this section were verified numerically for gravitational perturbations in the wormhole scenario. In particular, Fig.~\ref{fig:excitationwormhole} (upper left panel) displays the imaginary part of the first long-lived modes with $l=2$ as a function of the throat location $r_{\rm throat}$, while the upper right panel shows the corresponding derivative $d A^{\rm in}_{lm\omega}/d\omega$. Equation~\eqref{eqproduct} holds in the regime $|{\omega^I_n}| \ll |{\omega^R_n}| \ll M^{-1}$, where our numerical computation is reliable; this agreement is clearly visible in the lower left panel of Fig.~\ref{fig:excitationwormhole}. Outside this regime, the approximation breaks down. The lower right panel shows the corresponding values of the QNEFs $B_n$.

Employing Eqs.~\eqref{BnomegaI} and \eqref{eq:wormholeim}, one finds
\begin{equation} \label{eq:Bnscaling}
\left| \frac{B_n}{B_0} \right| \sim n^{2\ell+1},
\end{equation}
indicating that the excitation factor increases rapidly with the overtone number.
Although this scaling is derived in the specific low-frequency regime, it suggests
a general trend for cavity modes: higher overtones tend to possess larger QNEFs.

{\bf ECO case.} Using again a transfer matrix approach, Ref.~\cite{Rosato:2025byu} demonstrates that, for generic ECOs with surface reflectivity $R_{\rm ECO}$, the ingoing wave amplitude $A^{\rm in}_{lm\omega}$ can be computed as
\begin{equation}\label{eqAinECO}
   A^{\rm in}_{lm\omega}=\alpha_{lm\omega}^{\rm in}e^{-2i\omega L}+\alpha_{lm\omega}^{\rm out,*}R_{\rm ECO}e^{2i\omega L}\,,
\end{equation}
implying
\begin{align}
    \omega_n& = {\pi n \over L} +\frac{i}{4L} \log\left( \frac{\alpha^{\rm out}_{lm\omega_n}}{\alpha^{\rm in}_{lm \omega_n}} R_{\rm ECO} \right)\sim\notag\\&\sim{\pi n \over L}-\frac{i \beta_{ls}}{2L} \left(2M{\omega^R_n}\right)^{2l+2}+{i \over 4L}\log\left(|R_{\rm ECO}|\right)\,.
\end{align}
The structure of the imaginary part in the above expression is particularly
noteworthy. For reflecting ECOs, the damping rate of the modes is controlled by a
power-law dependence on the real part of the frequency.
From a spectral perspective, this places ECO resonances close to obstacle-type
scattering problems~\cite{Zworski:1999}, since dissipation is governed by leakage
through the effective potential barrier rather than by absorption at an horizon.

Hence, we obtain
\begin{equation}
 \frac{d A^{\rm in}_{lm\omega}}{d\omega} \sim 4iL\, \alpha^{\rm out,*}_{lm \omega} R_{\rm ECO} e^{i \omega L}\,,
\end{equation}
which corresponds to
\begin{equation}\label{dAdomegaECO}
 \left|\frac{d A^{\rm in}_{lm\omega}}{d\omega}\right| \sim \frac{2L}{\sqrt{\beta_{ls}}} \left(2M{\omega^R_n}\right)^{-l-1} |R_{\rm ECO}|\,.
\end{equation}
{We also need to compute $A^{\rm out}_{lm \omega}$, that for generic ECOs reads \cite{Rosato:2025byu}
\begin{equation}
    A^{\rm out}_{lm \omega}=\alpha^{\rm out}_{lm\omega}e^{-2 i \omega L}+ \left({1 \over \alpha^{\rm in}_{lm\omega}}+{\alpha^{\rm out}_{lm\omega}\alpha^{\rm out,*}_{lm\omega} \over \alpha^{\rm in}_{lm\omega}}\right)R_{\rm ECO} e^{2 i \omega L}\,.
\end{equation}
By substituting it into Eq.~\eqref{eqQNMS} evaluated in the ECO case, we obtain
\begin{equation}
    |A^{\rm out}_{lm \omega}|=\left|{R_{\rm ECO} \over \alpha^{\rm in}_{lm\omega}}e^{2 i \omega L}\right| \sim { 2 \sqrt{\beta_{ls}} |R_{\rm ECO}|\over \left(2M\omega_n^R\right)^{-l-1}}.
\end{equation}
This yields again a suppressed excitation factor:
\begin{equation}\label{dAdomegaECO}
|B_n| \sim
\beta_{ls}\frac{M}{L} \left(2M{\omega^R_n}\right)^{2l+1} \,.
\end{equation}
Indeed, long-lived modes arise again in the regime \({\omega^I_n} \ll {\omega^R_n} \ll M^{-1}\), which leads to small QNEFs. The above discussion demonstrates that the weak excitation of long-lived modes is a universal feature of ECOs as well as wormholes.

The results presented in this section have been numerically validated for gravitational perturbations in the context of different ECO models, in particular a Boltzmann reflectivity ECO~\cite{Oshita:2019sat} and configurations with constant reflectivity. Specifically, Fig.~\ref{fig:excitationECO} shows the behavior of the imaginary part of the first long-lived modes for \( l = 2 \) as a function of the object's radius \( r_0 \) in the upper panel. In the middle panel the corresponding \(\mathrm{d}A^{\rm in}_{{l} m\omega}/\mathrm{d}\omega\) are showed, while in the bottom panel we show the corresponding \(B_n\).
\begin{figure}[ht]
    \centering    \includegraphics[width=\linewidth]{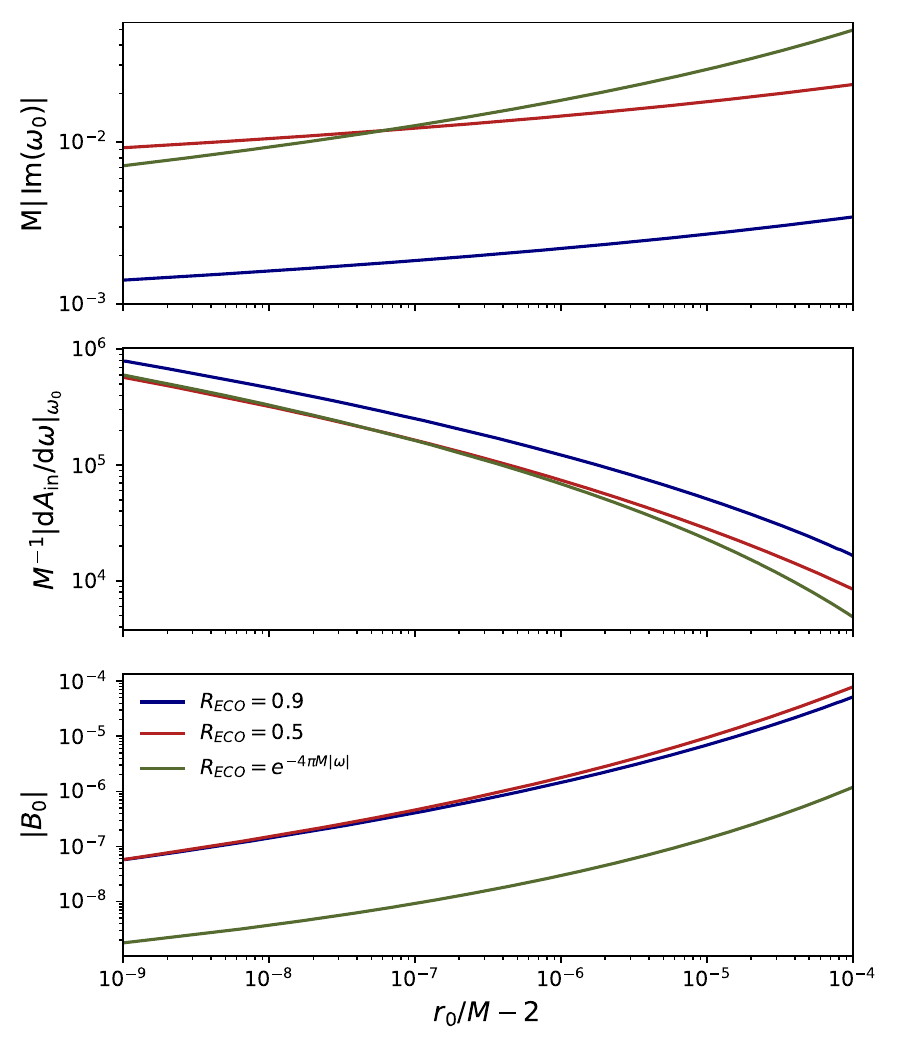}
    \caption{All panels refer to the $l=2$ mode of a gravitational perturbation for different ECO scenarios.
 \emph{Top panel:} imaginary part of the fundamental long-lived quasinormal mode for gravitational perturbations as a function of the object's radius \( r_0 \), for three different ECO reflectivities: constant reflectivity with \( R_{\rm ECO} = 0.9 \), \( R_{\rm ECO} = 0.5 \), and a Boltzmann reflectivity profile.  \emph{Middle panel:} corresponding behavior of the absolute value of the $dA^{\rm in}/d\omega$ for the same modes and models. Again, for long-lived modes the derivative \( \dv{A^{\rm in}_0}{\omega} \) is large, which implies that the corresponding modes are only weakly excited. \emph{Bottom panel:} the corresponding values of the QNEFs $B_0$ for the cases under consideration.
}
    \label{fig:excitationECO}
\end{figure} 

We notice that the scaling relation~\eqref{eq:Bnscaling}, derived for
wormholes, continues to hold also in the more general ECO case.

\section{Echo reconstruction from the excitation factors of ECOs} \label{sec:echoreconstruction}
In this section we will show how to reconstruct the ECO response from the QNEFs. To this purpose, it is instructive to compute the Fourier transform of the reflectivity, namely  
\begin{equation}\label{FTrefl}
    \mathcal{R}_{lm}(t)={1 \over 2\pi}\int_{-\infty}^{+\infty}d\omega R_{lm} e^{-{i} \omega t}\,,
\end{equation}
as this quantity captures the spectral content of the reflected signal.  
In particular, following Ref.~\cite{Rosato:2024arw}, the reflectivity can be evaluated either directly or by analyzing the analytic structure of the ingoing amplitude in the complex frequency plane~\cite{Leaver:1986gd}. The dominant contribution comes from the simple poles located at the QNM frequencies, corresponding to the complex roots of the ingoing amplitude.  
Applying the residue theorem, one obtains 
\begin{equation}
   \mathcal{R}_{lm}(t) \approx 4 {\rm Re}\left[ \sum_{n=1}^{n_{\rm max}}  \omega_{n}B_n e^{-i\omega_{n} t}\right]\,, \label{QNMdecomposition}
\end{equation}
where the dependence of $\omega_n$ and $B_n$ on $(l,m)$ is left implicit.
Within this framework, it becomes clear that the magnitude of the QNEFs, $B_n$, plays a fundamental role in understanding the echo signal. As previously discussed, the QNM spectrum of horizonless ultracompact objects is characterized by a set of modes with small imaginary part, arising because of the presence of an effective cavity in the potential. Among these cavity modes, long-lived modes correspond to a subset with particularly small imaginary part. However, all cavity modes feature small imaginary parts (smaller than those of standard BH QNMs, for reference). As shown by the analytical results of the previous section, this implies that cavity modes are generally weakly excited (e.g., when compared to the QNMs of a standard BH). 
Nevertheless, their small imaginary parts also make them only weakly damped over time, so that they decay appreciably only on long timescales. 
The excitation coefficients $B_n$ determine which of these modes dominate the signal at different stages of the evolution. 

Importantly, the early ringdown of wormholes and ECOs is identical to that of a standard BH with the same mass at early times~\cite{Cardoso:2016rao,Konoplya:2016hmd,Cardoso:2016oxy}. This behavior can be understood from a simple physical perspective.
Wormholes and ECOs effectively introduce a cavity of length $L$ in the scattering problem.
As a result, modifications to the waveform associated with reflections inside the
cavity can only affect the signal after an amount of time of order $2L$ has passed. At earlier times, the dynamics is entirely governed by the same effective
potential as in the BH case, implying that the early ringdown signal is
identical to that of a standard BH with the same mass.

In  Ref.~\cite{Rosato:2025byu}, it was shown that the reflectivity in the frequency domain exhibits oscillatory features \footnote{These features also appear in the gravitational–wave spectral amplitude,
where the reflectivity is imprinted \cite{Rosato:2025byu}, and consequently
in the energy spectrum \cite{Nair:2025anr}.
}. 
The same work demonstrated that, by isolating the high-frequency oscillations of the reflectivity and performing a Fourier transform, one can reconstruct the late-time echoes in the time domain. 
This was achieved by \emph{gluing} the reflectivity of a non-rotating wormhole to that of a reference Schwarzschild BH with the same mass. 
The matching was performed at a reference frequency $\omega_{\rm ref}=\sqrt{V_{\rm max}}$ (where $V_{\max}$ is the peak of the effective potential for the perturbation under consideration): at low frequencies, the contribution corresponds to that of a fictitious BH with the same mass as the wormhole, whereas at high frequencies the exact wormhole reflectivity is used. 
This construction enforces that the prompt ringdown of wormholes and ECOs coincides with that of a standard BH, while the first echoes arise predominantly from high-frequency features of the reflectivity.

Here, we extend this methodology to QNMs, proposing a framework to reconstruct the time-domain signal of horizonless ultracompact objects. In principle, by retaining a sufficiently large set of ECO oscillation modes, one could reproduce the entire ringdown, echoes included \footnote{A related discussion in a different but conceptually equivalent framework is given in Ref.~\cite{Oshita:2025ibu}.}. In practice, however, this would require a very large number of modes (typically tens to hundreds)~\cite{Oshita:2025ibu} rendering such an approach phenomenologically impractical. Therefore, along the lines of~\cite{Rosato:2025byu}, here we show that it is possible to proceed by modeling the early ringdown phase using standard BH QNMs, and the echo signal using only the oscillation frequencies of the cavity. In practice, we use the following reconstruction
\begin{multline}
   \mathcal{R}_{lm}(t) \approx \theta(2L-t)\,4 {\rm Re}\left[ \sum_{n=1}^{n_{\rm max}}  \omega^{\rm BH}_{n}B^{\rm BH}_n e^{-i\omega_{n}^{\rm BH} t}\right]\\
   +\theta(t-2L)\,4 {\rm Re}\left[ \sum_{n=1}^{n_{\rm max}}  \omega^{\rm cavity}_{n}B^{\rm\,cavity}_n e^{-i\omega_{n}^{\rm cavity} t}\right] \,,
\label{QNMdecompositionWORMHOLE}
\end{multline}
where the suffix indicates if the considered QNMs and QNEFs are associated to either BH or cavity modes, and $L$ is the cavity length. The resulting time-domain signal reproduces the  waveform, as shown in Fig.~\ref{fig:timedomainRefl} for the wormhole scenario, with high accuracy.
We stress that this framework is purely phenomenological, since BH QNMs are \emph{not} part of the ECO spectrum.
\begin{figure}[h]
    \centering    \includegraphics[width=\linewidth]{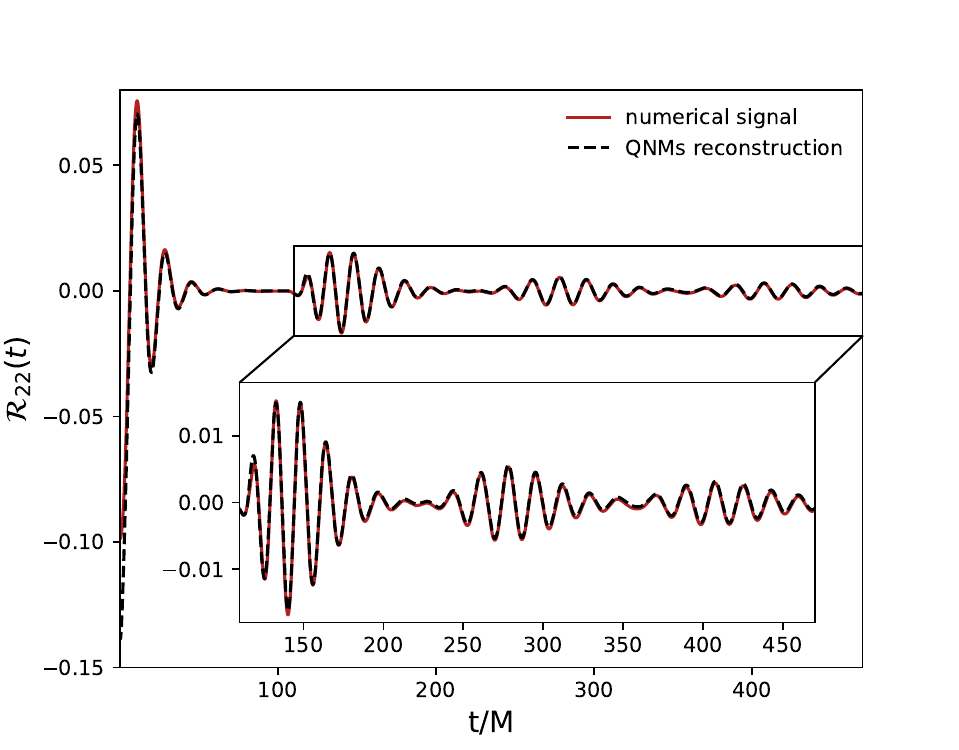}
    \caption{
Fourier transform of the reflectivity for a wormhole with throat location $r_{\rm throat} = (2 + 10^{-6})M$.  
To reconstruct the time-domain signal, we employ Eq.~\eqref{QNMdecompositionWORMHOLE}, with  $n^{\rm BH}_{\rm max} = 7$ for BH QNMs and $n^{\rm cavity}_{\rm max} = 12$ for cavity modes. Here the cavity length corresponds to $L \sim 54 M$. The resulting waveform accurately reproduces the oscillatory structure of the signal. 
The matching at $t = 2L$ is sufficiently smooth, since for sufficiently compact objects (i.e., large $L$) the initial oscillations have already fully decayed by that time, while the beat-like oscillations associated with the cavity modes vanish at $t = 2L$.
}   
\label{fig:timedomainRefl}
\end{figure}

Our aim now is to gain a deeper understanding of the structure of echoes. 
In Fig.~\ref{fig:timedomainRefl} we employed a large number of modes in Eq.~\eqref{QNMdecompositionWORMHOLE} to reproduce the time-domain signal. 
It is worth emphasizing that, in this theoretical analysis, no fitting procedure is involved: we simply compare a numerical time-domain result with its analytical counterpart given by Eq.~\eqref{QNMdecompositionWORMHOLE}. 
However, in a realistic data-analysis context, where one attempts to fit observational data using theoretical QNMs, employing too many modes is not viable, as it may lead to overfitting issues \cite{Baibhav:2023clw}.
Later on, we will show that such a large number of modes is in fact unnecessary.

Up to now, we have discussed that the wormhole/ECO signal can be interpreted as consisting of two distinct stages:
\begin{itemize}
  \item \textbf{Early ringdown.} As phenomenologically shown, this stage can be accurately reproduced using the modes of a fictitious BH with the same mass as the wormhole or ECO. As known from the BH case, only a few modes are sufficient to obtain a good reconstruction of the signal.
  \item \textbf{Echoes stage.} This phase is seeded by the cavity modes, which manifest as beat-like oscillations resulting from the interference between the trapped modes within the effective cavity.
\end{itemize}
However, the contribution of cavity modes to echoes is not the same at every stage. In particular, one must understand {\it how} cavity modes contribute. Indeed, the excitation of these modes scales with $\omega_n^I$, as previously shown. 

Echoes are typically associated with the long-lived modes, which however constitute only a limited portion of the cavity-mode spectrum. Here, the first key point we wish to emphasize is that the echoes are not produced \emph{solely} by the long-lived modes; rather, they are generated by the entire set of cavity modes, which is larger than the subset of long-lived modes. In particular, as we have shown, the longer-lived a mode is (i.e. the smaller its imaginary part in absolute value), the more strongly its excitation factor is suppressed. For this reason, we do not expect the long-lived modes to contribute significantly to the first echoes, which have the largest amplitudes.

In Fig.~\ref{fig:mode_convergence} we display a representative portion of the spectrum of cavity modes for a specific wormhole configuration with $r_{\rm throat}=2M(1+10^{-7})$. 
To disentangle the role of the long-lived modes, we introduce a reference frequency corresponding to $\omega_{\rm ref}\sim \sqrt{V_{\rm max}}$, following Ref.~\cite{Rosato:2025byu}, where $V_{\max}$ is the peak of the effective potential for the perturbation under consideration (axial gravitational ones, in this example). 
Long-lived modes lie below this reference frequency, as shown in the left panel of the figure. In the right panel, the black solid line represents the echo portion of the numerical time-domain reflectivity.  We then attempt to reconstruct the signal by (i) including only the modes with real parts below the reference frequency (light-blue dashed line), and (ii) including all cavity modes (orange dashed line). As can be deduced from Fig.~\ref{fig:mode_convergence}, the modes with extremely small imaginary parts contribute predominantly to \emph{very}-late-time echoes, but not to the first echoes of the signal. Indeed, in the figure the light-blue dashed line reproduces the last echo more accurately than the first two, though not perfectly.

From a phenomenological perspective, however, the first echoes are the most relevant ones. 
These are instead governed by modes with larger imaginary parts, which typically also have larger real parts compared to those of the long-lived branch, namely $\omega_{R}>\omega_{R}^{\text{BH}}\sim\sqrt{V_{\text{max}}}$. 
This behavior is fully consistent with our previous finding that early echoes are driven by the high-frequency features of the system~\cite{Rosato:2025byu}. 
As illustrated in the right panel of Fig.~\ref{fig:mode_convergence}, when high-frequency modes are included, the echoes are faithfully reconstructed. If they are excluded, the echo structure is lost.
\begin{figure*}[ht]
  \centering
\includegraphics[width=0.9\linewidth]{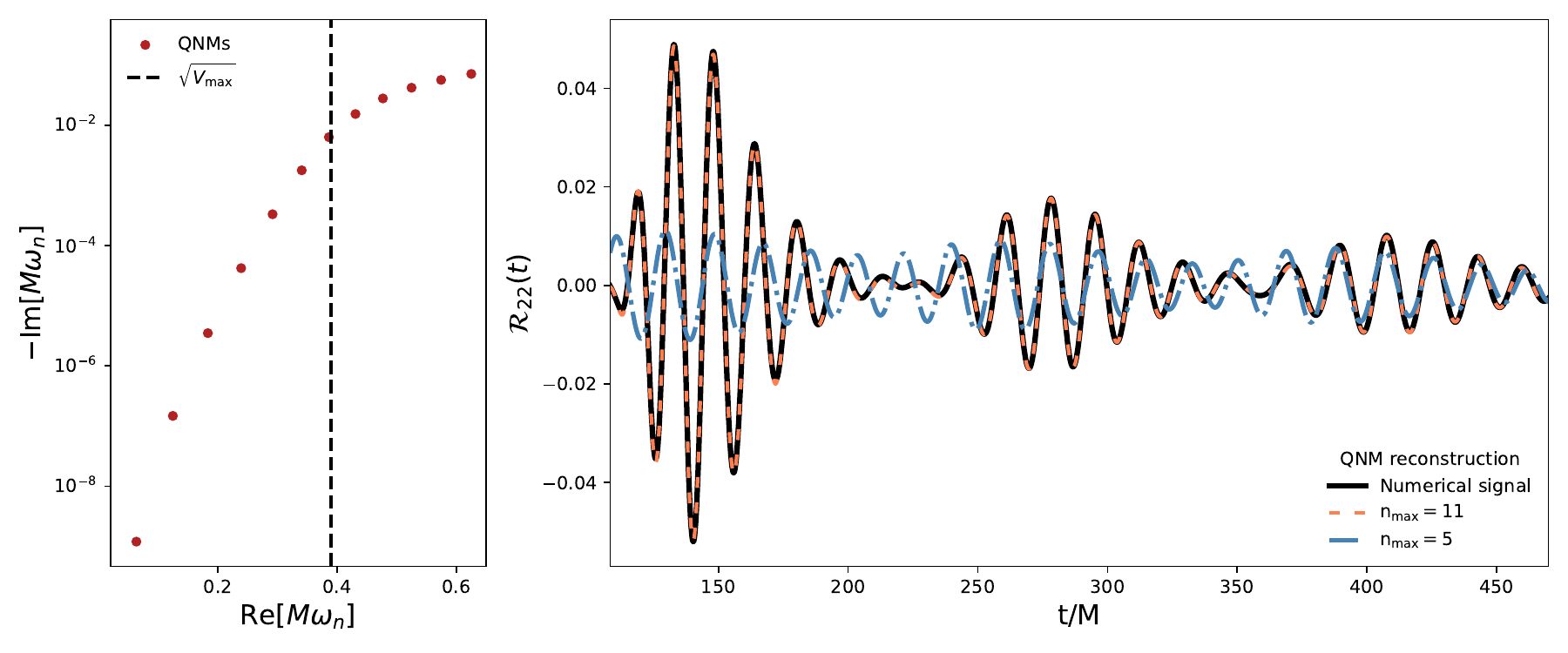}
  \caption{{\it Mode content and echo reconstruction} for a wormhole with throat location at $r_{\rm throat}=(2+10^{-6})M$ 
  \emph{Left:} representative portion of the cavity-mode spectrum. The longest-lived modes do not dominate the echo onset because their excitation factors are highly suppressed. 
  \emph{Right:} comparison between the numerical echo signal and QNM-based reconstructions with/without high-frequency modes (defined as those with ${\omega^R_n}>\sqrt{V_{\max}}$). Including high-frequency modes yields an accurate reconstruction of the first echoes; excluding them reproduces only very-late-time echoes.}
  \label{fig:mode_convergence}
\end{figure*}

{\bf Minimal waveform for BH ringdown+echoes.}
We have therefore demonstrated that high-frequency cavity modes dominate the formation of the first echoes in the ringdown of horizonless ultracompact objects. 
However, in the previous reconstruction we still employed a large number of modes. 
The next question we address is: \emph{how many of these high-frequency cavity modes are actually needed to reconstruct the echoes?} 
Our second key result is that only a few such modes are sufficient. In general, only a small number of modes (between $2$ and $4$) is required to reconstruct a single echo. In Fig.~\ref{fig:mode_convergence_bis}, we show the case of an echo in the wormhole signal with throat location at $r_{\rm throat}=2.0001M$. Already two modes (dashed blue line) provide a qualitatively good reconstruction of the numerical signal (black solid line), while four modes (dashed orange line) further improve the agreement, leading to a satisfactory reproduction of the entire echo structure.
\begin{figure}[h]
  \centering  \includegraphics[width=0.85\linewidth]{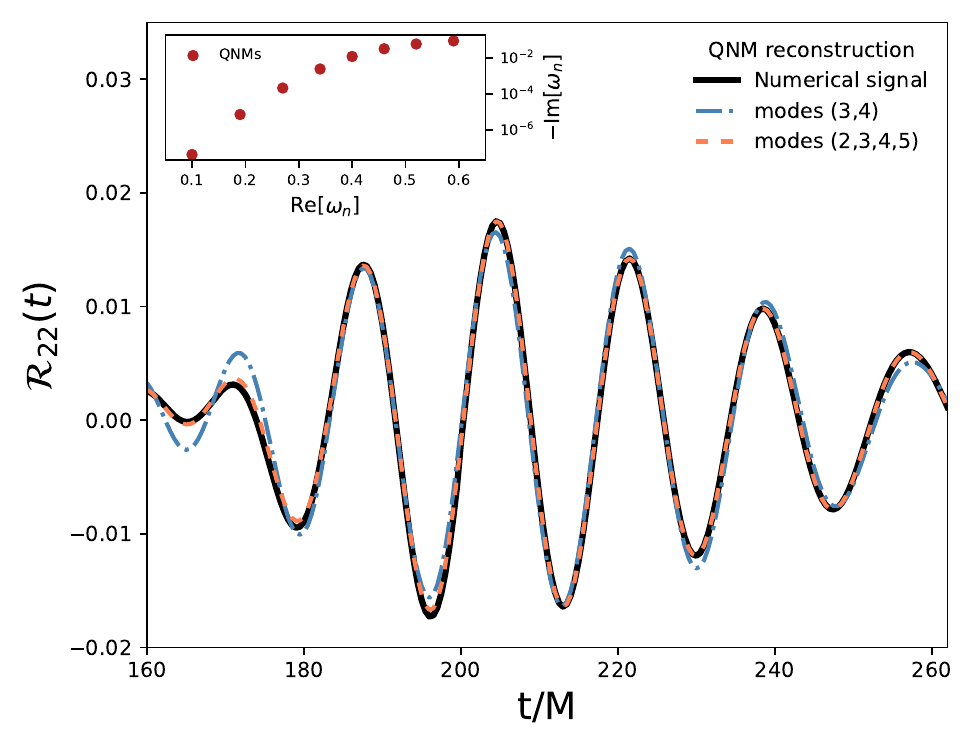}
  \caption{Reconstruction of a wormhole echo with throat location $r_{\rm throat}=2.00001M$. The inset in the upper left shows a representative set of QNMs. The numerical signal is shown as a black solid line, focusing on the second echoes in the interval $4L<t<6L$. The dashed blue line corresponds to a reconstruction with two modes, specifically the fourth and fifth ones, i.e. the first two modes above the threshold $\sqrt{V_{\rm max}}$, which already provide a good agreement. The dashed orange line shows the case with four modes, yielding an even better match. This demonstrates that only a small number of modes is sufficient to reproduce the echo.
}
  \label{fig:mode_convergence_bis}
\end{figure}

Based on the above findings, we propose a phenomenological ringdown waveform 
\begin{align}\label{template}
    h(t)=&\theta(2L-t)\sum_{{l} mn} A_{{l} mn}\cos\left(\omega_{lmn}^Rt+\phi_{{l} mn}\right)
    e^{\omega_{lmn}^I t}\nonumber\\
    &+\theta(t-2L)\sum_{{l} mn} \hat A_{{l} mn}\cos\left(\hat\omega_{lmn}^Rt+\hat \phi_{{l} mn}\right)e^{\hat\omega_{lmn}^I t}\,,
\end{align}
where $A_{{l} mn}$ and $\phi_{{l} mn}$  denote the amplitude and phase of the $({l},m,n)$ BH mode, with $({l},m,n)$ being the angular, azimuthal, and overtone number, respectively. The corresponding hatted quantities refer instead to cavity modes. 

The first line of Eq.~\eqref{template} is the usual BH ringdown waveform truncated at $t=2L$, while the second line implements the echo reconstruction with the cavity modes previously discussed. This phenomenological model is similar in spirit to the QNM superposition studied in Ref.~\cite{Maselli:2017tfq}, but informed by the cavity modes.

Compared to a standard ringdown waveform, the model above has $4N$ extra parameters when $N$ cavity modes are considered. Indeed, each mode is described by the amplitude, phase, frequency, and damping time; one of the phases must be fixed to require continuity of the waveform at $t=2L$, but this is compensated by the extra $L$ parameter. For example, considering two cavity modes only, one would need to include 8 extra parameters. A given ECO model would fix all the cavity QNMs in terms of $\epsilon$, effectively reducing the number of extra parameters to $2N+1$.

\section{Stability of long-lived modes}\label{stability}
It is well established that BH QNMs are highly sensitive to small perturbations of the system~\cite{Nollert:1996rf,Daghigh:2020jyk,Jaramillo:2020tuu,Jaramillo:2021tmt,Warnick:2024usx}, whether these affect the background geometry or the boundary conditions. This strong sensitivity implies that the QNM spectrum of a BH can be significantly distorted in the presence of environmental effects~\cite{Barausse:2014tra,Barausse:2014pra,Cheung:2021bol,Berti:2022xfj} or any structure near the horizon~\cite{Cardoso:2016rao,Cardoso:2016oxy,Cardoso:2017cqb,Abedi:2020ujo}. Nevertheless, the early-time ringdown signal in the time domain is generally much less susceptible to such modifications~\cite{Cardoso:2016rao,Cardoso:2016oxy,Cardoso:2017cqb,Mirbabayi:2018mdm,Berti:2022xfj,Kyutoku:2022gbr}.

Here we given an analytical argument showing that, for ECOs and wormholes, the long-lived QNMs are remarkably robust against localized perturbations of small amplitude, as recently found numerically~\cite{Destounis:2025dck}. In particular, modes with smaller imaginary parts tend to be more stable. To demonstrate this, let us consider the wave equation~\eqref{Regge-Zerilli}. We model the perturbation of the potential as a localized deformation of amplitude \( \epsilon V_{\rm bump}(r_*) \), with \( \epsilon \ll 1 \). Provided that the transfer matrix approximation holds~\cite{Ianniccari:2024ysv}, and placing the bump to the right of the unperturbed potential \( V_l(r_*) \) (without loss of generality), the scattering problem can be described as~\cite{Ianniccari:2024ysv}:
\begin{equation}
    \psi(r_* = +\infty) = \mathcal{M}_{\rm bump} \, U(c) \, \mathcal{M}_0 \, \psi(r_* = -\infty)\,,
\end{equation}
where \( \mathcal{M}_0 \) and \( \mathcal{M}_{\rm bump} \) are the transfer matrices corresponding to the background potential and the bump respectively, \( U(c) \) is a translation operator by a distance \( c \), representing the position of the bump. 

The total transmission coefficient of the system is given by
\begin{align} \label{totalT}
    \frac{1}{T_{\rm tot}} &= -\frac{R'_{\rm bump} R_0}{T_{\rm bump} T_0} e^{i \omega c} + \frac{1}{T_{\rm bump} T_0} e^{-i \omega c} = \notag\\
   &  e^{-i \omega c} \left( A^{\rm in}_0 A^{\rm in}_{\rm bump} - A^{\rm out}_0 A'^{\rm out}_{\rm bump} e^{2i \omega c} \right)\,,
\end{align}
and QNMs correspond to the roots of Eq.~\eqref{totalT}.

{ \bf Perturbative analysis around a long-lived mode.} 
We now focus on a long-lived QNM of the unperturbed system, characterized by a frequency \( \omega_n \) with small imaginary part. For frequencies \( M\omega > \epsilon \), the outgoing amplitude associated with the bump is strongly suppressed, while the ingoing amplitude approaches unity. Specifically, we can write
\begin{equation}
    A'^{\rm out}_{\rm bump}\sim \epsilon \, a(\omega)\,, \qquad 
    A^{\rm in}_{\rm bump} \sim e^{i\delta} - \epsilon \, b(\omega)\,,
\end{equation}
where \( a(\omega) \) and \( b(\omega) \) are finite complex functions of the frequency, and \( \delta \in \mathbb{R} \) is a phase shift.

Near the unperturbed QNM frequency \( \omega_n \), we can expand the ingoing amplitude of the background system as
\begin{equation}
    A^{\rm in}_0(\omega) \simeq \left. \frac{d A^{\rm in}_0}{d\omega} \right|_{\omega_n} (\omega - \omega_n)\,.
\end{equation}
Substituting into Eq.~\eqref{totalT}, and normalizing \( A^{\rm out}_0 = 1 \), we obtain
\begin{equation}
    \left. \frac{d A^{\rm in}_0}{d\omega} \right|_{\omega_n} (\omega - \omega_n) \left( e^{i\delta} - \epsilon \, b(\omega) \right) - \epsilon \, a(\omega) e^{2i\omega c} = 0\,.
\end{equation}
This equation can be solved perturbatively by setting \( \omega = \omega_n + \Delta\omega_n \), which yields
\begin{equation} \label{eqDeltaomega}
    \Delta\omega_n = 2\epsilon \omega_n e^{-i\delta+2i\omega_n c} a(\omega_n) B_n \,,
\end{equation}

where, interestingly, the QNEF $B_n$ also appear naturally in this context.

{\bf Implications for mode stability.} Eq.~\eqref{eqDeltaomega} shows that the frequency shift \( \Delta\omega_n \) remains of order \( \mathcal{O}(\epsilon) \) as long as the condition \( 2c < 1/|{\rm Im}(\omega_n)| \) is satisfied. However, when \( 2c > 1/|{\rm Im}(\omega_n)| \), the correction grows exponentially with \( c \), signaling a potential instability. Therefore, modes with a larger imaginary part (i.e., shorter-lived) are more susceptible to instability, while long-lived modes exhibit enhanced stability under local perturbations.

Furthermore, $\Delta \omega_n$ for long-lived QNMs is further suppressed due to the small value of the QNEF \(  B_n \), which scales as \(\sim|{\omega^I_n}| \), as previously discussed.

We stress that this behavior reflects a well-known feature of non-self-adjoint
spectral problems.
In the spectral-analysis literature, eigenvalues with larger absolute values of
the imaginary part are known to exhibit enhanced sensitivity to perturbations
\cite{Warnick:2024usx,Jaramillo:2020tuu, Hitrik:2025}.

{\bf Numerical verification.} To support the analytical findings above, we will now numerically investigate the effect of small perturbations by modifying the effective potential \( V_l \) as follows 
\begin{equation}\label{bump}
    V_{\rm eff}(r_*) = V_l(r_*) + {\epsilon \over M^2} \,\text{Sech}(r_*-c)^2\,.
\end{equation}

In the case of a symmetric wormhole geometry, we preserve the mirror symmetry by applying the same perturbation in both universes, placing it symmetrically with respect to the throat location \( r_0 \), as done in the standard configuration. 

Figure~\ref{fig:responsemodes} illustrates the response of the system to the localized perturbation \eqref{bump}, comparing three representative QNMs: a standard (rapidly damped) mode, a long-lived mode, and an intermediate (cavity) mode. The left panel shows the relative change of ${\omega^R_n}$ as a function of the bump location $c/M$, while the right panel reports the relative change of ${\omega^I_n}$. For the data shown we set the bump amplitude to $\epsilon=10^{-4}$ (the magnitude of the perturbation can be estimated as $\epsilon/(M^2 V_{\rm max})\simeq6.64\times10^{-4}$). The plotted complex frequencies are: standard mode $M\omega_{\rm standard}=0.590-0.0894\,i$ (blue) , long-lived mode $M\omega_{\rm long}=0.189-7.086\times10^{-6}\,i$ (red), and intermediate (cavity) mode $M\omega_{\rm int}=0.401-0.012\,i$ (green). The three modes display markedly different behavior: the standard mode is strongly destabilized, with relative shifts that rapidly exceed the perturbation scale $\mathcal{O}(\epsilon)$. By contrast, the long-lived mode is essentially unaffected, its real part shifting by orders of magnitude less than $\epsilon$ and its imaginary part changing at most by a factor of order unity. The intermediate (cavity) mode exhibits an intermediate response. Consistently with Eq.~\eqref{eqDeltaomega}, its relative variation begins to exceed the perturbative magnitude for $c\sim 1/|\Im\omega|_{\rm int}\simeq83M$, and the exponential growth predicted by Eq.~\eqref{eqDeltaomega} is visible over the displayed range of $c/M$.

\begin{figure*}[ht]
    \centering
    \includegraphics[width=0.95\textwidth]{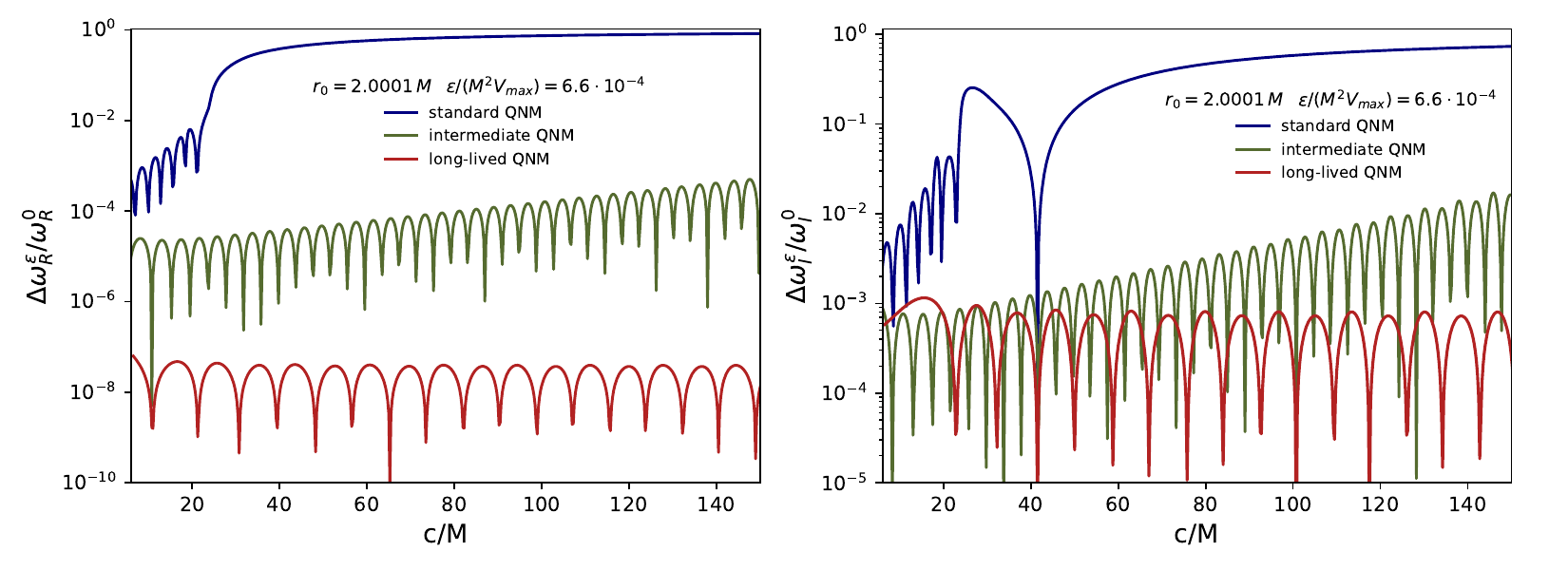}
    \caption{Response of three QNMs to the localized perturbation in Eq.~\eqref{bump} with $\epsilon=10^{-4}$ for a wormhole with throat \(r_0=2.0001M\). \emph{Left:} relative change of \(\Re\omega\) as a function of \(c/M\). \emph{Right:} relative change of \(\Im\omega\) as a function of \(c/M\). The curves correspond to: a standard (rapidly damped) QNM (blue solid line) whose imaginary part is comparable with the one of Schwarzschild fundamental mode, \(M\omega_{\rm standard}=0.590-0.0894\,i\); a long-lived mode (red solid line) with \(M\omega_{\rm long}=0.189-7.086\times10^{-6}\,i\); an intermediate cavity mode (green solid line), with \(M\omega_{\rm int}=0.401-0.012\,i\), whose damping rate lies between that of the Schwarzschild fundamental mode and that of the long-lived mode. The standard mode is strongly destabilized, whereas the long-lived mode remains essentially unaffected. The intermediate mode shows intermediate behavior. See the main text for quantitative details.}
  \label{fig:responsemodes}
\end{figure*} 
\section{Stability of Greybody factors} \label{sec:stabilityGF}
Recent work has investigated the stability of GFs under small perturbations of the system, as well as their connection with observable quantities~\cite{Rosato:2024arw,Oshita:2023cjz,Oshita:2024fzf,Okabayashi:2024qbz}. We emphasize that GFs are global scattering observables and, as such, their
stability is not tied to the spectral stability of individual quasinormal modes.
In particular, their robustness persists even in regimes where the underlying QNM
spectrum is known to be spectrally unstable \cite{Rosato:2024arw} \footnote{See \cite{Torres:2023nqg} for a related discussion.}.  

 Here, we consider the modification introduced in Eq.~\eqref{bump} and numerically
verify that greybody-factor stability holds also in the wormhole and ECO scenarios.
Recalling that the greybody factors are functions of frequency, we quantify the
corresponding variation through the relative difference

\begin{equation}\label{Deltal}
    \mathcal{G}_{lm} = 
    \frac{ \int_0^{\infty} 
    \left| \Gamma^\epsilon_{lm} (\omega, c) - \Gamma_{lm} (\omega) \right| \, d\omega }
    { \int_0^{\infty} \Gamma_{lm} (\omega) \, d\omega }\,,
\end{equation}
where the unperturbed greybody factors \( \Gamma_{lm}(\omega) \) and the perturbed ones \( \Gamma^\epsilon_{lm}(\omega, c) \) are computed via direct numerical integration, employing high-order analytic series expansions to ensure high precision~\cite{Pani:2013pma,Brito:2015oca}.

\begin{figure*}[ht]
    \centering   
    \includegraphics[width=0.95\textwidth]{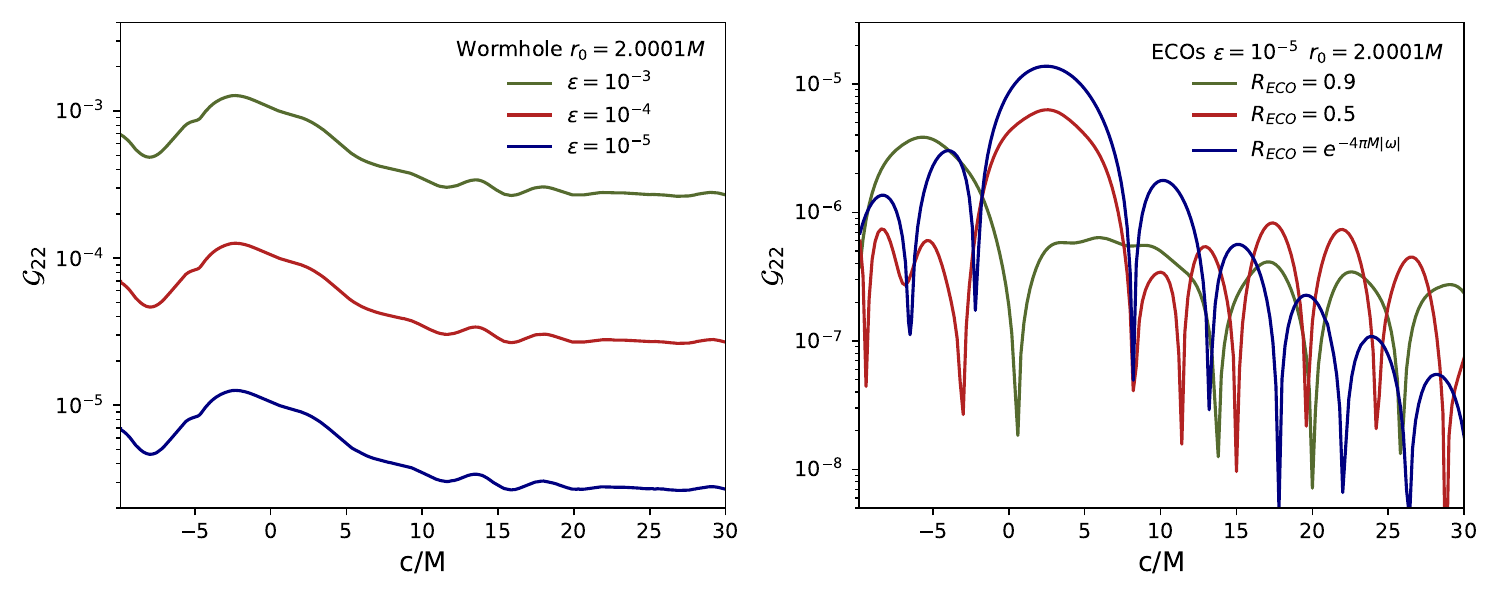}
    \caption{
    Relative variation \( \mathcal{G}_{lm} \) of the greybody factors under a localized perturbation of the effective potential, as introduced in Eq.~\eqref{bump}, shown as a function of its location \( c/M \) and defined in Eq.~\eqref{Deltal}. 
\emph{Left panel:} wormhole case for different values of the perturbation amplitude \( \epsilon \). The variation remains consistently within the same order of magnitude as the perturbation. 
\emph{Right panel:} results for three different ECO models — constant reflectivity with \( R_{\rm ECO} = 0.9 \) (green), \( R_{\rm ECO} = 0.5 \) (red), and a Boltzmann reflectivity profile (blue). In all cases, the deviation does not exceed \( \mathcal{O}(\epsilon) \), confirming the stability of the GFs.
    }
    \label{fig:GF_variation}
\end{figure*}
This variation is illustrated in Fig.~\ref{fig:GF_variation}. 
The left panel shows the case of wormholes for different values of \( \epsilon \), where the variation never exceeds the order of magnitude of the corresponding perturbation. 
The right panel presents the results for different ECO configurations, namely: constant reflectivity with \( R_{\rm ECO} = 0.9 \) (green solid line), \( R_{\rm ECO} = 0.5 \) (red solid line), and a Boltzmann reflectivity profile (blue solid line). 
In all cases, the variation remains bounded within \( \mathcal{O}(\epsilon) \), confirming that both wormhole and ECO greybody factors are stable under small perturbations.
This behavior can be understood by noting that GFs probe the global
scattering properties of the effective potential rather than the detailed spectral
structure of the QNM eigenvalue problem.
While QNMs are associated with discrete, highly non-self-adjoint
eigenvalues and can therefore exhibit strong spectral instability, GFs
are defined through frequency-domain transmission coefficients, which remain smooth
functions of $\omega$ under localized perturbations.
As a result, the spectral instability of individual QNMs does not translate into an
instability of the corresponding GFs \cite{Rosato:2024arw}.
\footnote{See \cite{Torres:2023nqg} for a related discussion.}

\section{Conclusions}
In this work, we investigated the QNEFs and their connection to the ringdown of ultracompact horizonless objects. 
Using models such as Schwarzschild-like wormholes and partially absorbing ECOs, we systematically analyzed the excitation of their quasinormal spectra. 
Our results highlight several noteworthy features of QNEFs and their relevance to the physics of ultracompact objects.

First, we showed that long-lived QNMs (such as the cavity modes), which characterize the spectra of these systems, are only weakly excited during the ringdown. 
This observation has several implications: it explains why long-lived modes appear only at late times in the waveform, producing echoes. 
Since they are weakly excited, they do not affect the prompt ringdown, which remains dominated by standard BH (i.e., photon-sphere) modes. 
They become visible only after the standard QNMs have decayed, and ---being associated with cavity modes--- give rise to echoes as interference patterns. 
Moreover, the long-lived QNEFs scale as \( \sim |{\omega^I_n}| \), implying that higher-frequency modes (with larger absolute imaginary parts) contribute earlier in the signal, whereas the low-frequency part of the spectrum manifests only at very late times.

Based on this result, we proposed a practical ringdown waveform model (Eq.~\eqref{template}) based on a superposition of ordinary BH QNMs at early time and cavity modes at later times, which captures the whole complexity of the ringdown of horizonless ultracompact objects with a minimal set of waveform parameters. 

A second key result is that the combination of small QNEFs and small imaginary parts prevents long-lived modes from becoming spectrally unstable, as we demonstrated analytically.

Finally, we extended previous analyses of the stability of GFs to ECOs and wormholes, showing that GFs constitute robust observables even for horizonless ultracompact objects.

The present work opens several directions for future research. 
A natural extension is the generalization of our analysis to \emph{rotating} geometries, where both the structure of the effective potential and mode pattern become significantly richer. Another promising direction concerns the connection between QNEFs and the time-domain response in more realistic scenarios, e.g. in a particle infall experiment where the specification of the source allow us to compute the corresponding excitation coefficients. 
Finally, it would be interesting to devise search strategies based on our ringdown+echo waveform model, which captures the standard prompt BH ringdown and the first few echoes with a minimal set of parameters.
\section{Acknowledgments}
This work is partially supported by the MUR FIS2 Advanced Grant ET-NOW (CUP:~B53C25001080001) and by the INFN TEONGRAV initiative. SB thank PhD fellowship provided by IACS, as a part of this work was done when he was a PhD student there. SB thanks Alok Laddha and K. P Yogendran for helpful discussions. 
\appendix

\bibliography{biblio}

@article{Cardoso:2014sna,
    author = "Cardoso, Vitor and Crispino, Lu\'\i{}s C. B. and Macedo, Caio F. B. and Okawa, Hirotada and Pani, Paolo",
    title = "{Light rings as observational evidence for event horizons: long-lived modes, ergoregions and nonlinear instabilities of ultracompact objects}",
    eprint = "1406.5510",
    archivePrefix = "arXiv",
    primaryClass = "gr-qc",
    doi = "10.1103/PhysRevD.90.044069",
    journal = "Phys. Rev. D",
    volume = "90",
    number = "4",
    pages = "044069",
    year = "2014"
}

@article{Cardoso:2019rvt,
    author = "Cardoso, Vitor and Pani, Paolo",
    title = "{Testing the nature of dark compact objects: a status report}",
    eprint = "1904.05363",
    archivePrefix = "arXiv",
    primaryClass = "gr-qc",
    doi = "10.1007/s41114-019-0020-4",
    journal = "Living Rev. Rel.",
    volume = "22",
    number = "1",
    pages = "4",
    year = "2019"
}

@article{Maggio:2021ans,
    author = "Maggio, Elisa and Pani, Paolo and Raposo, Guilherme",
    title = "{Testing the nature of dark compact objects with gravitational waves}",
    eprint = "2105.06410",
    archivePrefix = "arXiv",
    primaryClass = "gr-qc",
    month = "5",
    year = "2021"
}

@article{Maggio:2017ivp,
    author = "Maggio, Elisa and Pani, Paolo and Ferrari, Valeria",
    title = "{Exotic Compact Objects and How to Quench their Ergoregion Instability}",
    eprint = "1703.03696",
    archivePrefix = "arXiv",
    primaryClass = "gr-qc",
    doi = "10.1103/PhysRevD.96.104047",
    journal = "Phys. Rev. D",
    volume = "96",
    number = "10",
    pages = "104047",
    year = "2017"
}

@article{Mark:2017dnq,
    author = "Mark, Zachary and Zimmerman, Aaron and Du, Song Ming and Chen, Yanbei",
    title = "{A recipe for echoes from exotic compact objects}",
    eprint = "1706.06155",
    archivePrefix = "arXiv",
    primaryClass = "gr-qc",
    reportNumber = "LIGO-P1700145",
    doi = "10.1103/PhysRevD.96.084002",
    journal = "Phys. Rev. D",
    volume = "96",
    number = "8",
    pages = "084002",
    year = "2017"
}

@article{Maggio:2020jml,
    author = "Maggio, Elisa and Buoninfante, Luca and Mazumdar, Anupam and Pani, Paolo",
    title = "{How does a dark compact object ringdown?}",
    eprint = "2006.14628",
    archivePrefix = "arXiv",
    primaryClass = "gr-qc",
    doi = "10.1103/PhysRevD.102.064053",
    journal = "Phys. Rev. D",
    volume = "102",
    number = "6",
    pages = "064053",
    year = "2020"
}

@article{Chakraborty:2022zlq,
    author = "Chakraborty, Sumanta and Maggio, Elisa and Mazumdar, Anupam and Pani, Paolo",
    title = "{Implications of the quantum nature of the black hole horizon on the gravitational-wave ringdown}",
    eprint = "2202.09111",
    archivePrefix = "arXiv",
    primaryClass = "gr-qc",
    doi = "10.1103/PhysRevD.106.024041",
    journal = "Phys. Rev. D",
    volume = "106",
    number = "2",
    pages = "024041",
    year = "2022"
}

@article{Rosato:2025byu,
    author = "Rosato, Romeo Felice and Biswas, Shauvik and Chakraborty, Sumanta and Pani, Paolo",
    title = "{Greybody factors, reflectionless scattering modes, and echoes of ultracompact horizonless objects}",
    eprint = "2501.16433",
    archivePrefix = "arXiv",
    primaryClass = "gr-qc",
    doi = "10.1103/PhysRevD.111.084051",
    journal = "Phys. Rev. D",
    volume = "111",
    number = "8",
    pages = "084051",
    year = "2025"
}

@book{Visser:1995cc,
    author = "Visser, Matt",
    title = "{Lorentzian wormholes: From Einstein to Hawking}",
    isbn = "978-1-56396-653-8",
    year = "1995"
}

@article{Maggio:2018ivz,
    author = "Maggio, Elisa and Cardoso, Vitor and Dolan, Sam R. and Pani, Paolo",
    title = "{Ergoregion instability of exotic compact objects: electromagnetic and gravitational perturbations and the role of absorption}",
    eprint = "1807.08840",
    archivePrefix = "arXiv",
    primaryClass = "gr-qc",
    doi = "10.1103/PhysRevD.99.064007",
    journal = "Phys. Rev. D",
    volume = "99",
    number = "6",
    pages = "064007",
    year = "2019"
}

@article{Rosato:2024arw,
    author = "Rosato, Romeo Felice and Destounis, Kyriakos and Pani, Paolo",
    title = "{Ringdown stability: Graybody factors as stable gravitational-wave observables}",
    eprint = "2406.01692",
    archivePrefix = "arXiv",
    primaryClass = "gr-qc",
    doi = "10.1103/PhysRevD.110.L121501",
    journal = "Phys. Rev. D",
    volume = "110",
    number = "12",
    pages = "L121501",
    year = "2024"
}

@article{Starobinskil:1974nkd,
    author = "Starobinskil, Alexei A. and Churilov, S. M.",
    title = "{Amplification of electromagnetic and gravitational waves scattered by a rotating ''black hole''}",
    journal = "Sov. Phys. JETP",
    volume = "65",
    number = "1",
    pages = "1--5",
    year = "1974"
}

@article{Cardoso:2016rao,
    author = "Cardoso, Vitor and Franzin, Edgardo and Pani, Paolo",
    title = "{Is the gravitational-wave ringdown a probe of the event horizon?}",
    eprint = "1602.07309",
    archivePrefix = "arXiv",
    primaryClass = "gr-qc",
    doi = "10.1103/PhysRevLett.116.171101",
    journal = "Phys. Rev. Lett.",
    volume = "116",
    number = "17",
    pages = "171101",
    year = "2016",
    note = "[Erratum: Phys.Rev.Lett. 117, 089902 (2016)]"
}

@article{Konoplya:2016hmd,
    author = "Konoplya, R. A. and Zhidenko, A.",
    title = "{Wormholes versus black holes: quasinormal ringing at early and late times}",
    eprint = "1606.00517",
    archivePrefix = "arXiv",
    primaryClass = "gr-qc",
    doi = "10.1088/1475-7516/2016/12/043",
    journal = "JCAP",
    volume = "12",
    pages = "043",
    year = "2016"
}

@article{Cardoso:2016oxy,
    author = "Cardoso, Vitor and Hopper, Seth and Macedo, Caio F. B. and Palenzuela, Carlos and Pani, Paolo",
    title = "{Gravitational-wave signatures of exotic compact objects and of quantum corrections at the horizon scale}",
    eprint = "1608.08637",
    archivePrefix = "arXiv",
    primaryClass = "gr-qc",
    doi = "10.1103/PhysRevD.94.084031",
    journal = "Phys. Rev. D",
    volume = "94",
    number = "8",
    pages = "084031",
    year = "2016"
}

@article{Maselli:2017tfq,
    author = {Maselli, Andrea and V{\"o}lkel, Sebastian H. and Kokkotas, Kostas D.},
    title = "{Parameter estimation of gravitational wave echoes from exotic compact objects}",
    eprint = "1708.02217",
    archivePrefix = "arXiv",
    primaryClass = "gr-qc",
    doi = "10.1103/PhysRevD.96.064045",
    journal = "Phys. Rev. D",
    volume = "96",
    number = "6",
    pages = "064045",
    year = "2017"
}

@article{Bueno_2018,
   title={Echoes of Kerr-like wormholes},
   volume={97},
   ISSN={2470-0029},
   url={http://dx.doi.org/10.1103/PhysRevD.97.024040},
   DOI={10.1103/physrevd.97.024040},
   number={2},
   journal={Physical Review D},
   publisher={American Physical Society (APS)},
   author={Bueno, Pablo and Cano, Pablo A. and Goelen, Frederik and Hertog, Thomas and Vercnocke, Bert},
   year={2018},
   month=jan }

@article{Oshita:2019sat,
    author = "Oshita, Naritaka and Wang, Qingwen and Afshordi, Niayesh",
    title = "{On Reflectivity of Quantum Black Hole Horizons}",
    eprint = "1905.00464",
    archivePrefix = "arXiv",
    primaryClass = "hep-th",
    doi = "10.1088/1475-7516/2020/04/016",
    journal = "JCAP",
    volume = "04",
    pages = "016",
    year = "2020"
}

@article{Nollert:1996rf,
    author = "Nollert, Hans-Peter",
    title = "{About the significance of quasinormal modes of black holes}",
    eprint = "gr-qc/9602032",
    archivePrefix = "arXiv",
    doi = "10.1103/PhysRevD.53.4397",
    journal = "Phys. Rev. D",
    volume = "53",
    pages = "4397--4402",
    year = "1996"
}

@article{Daghigh:2020jyk,
    author = "Daghigh, Ramin G. and Green, Michael D. and Morey, Jodin C.",
    title = "{Significance of Black Hole Quasinormal Modes: A Closer Look}",
    eprint = "2002.07251",
    archivePrefix = "arXiv",
    primaryClass = "gr-qc",
    doi = "10.1103/PhysRevD.101.104009",
    journal = "Phys. Rev. D",
    volume = "101",
    number = "10",
    pages = "104009",
    year = "2020"
}

@article{Jaramillo:2020tuu,
    author = "Jaramillo, Jos\'e Luis and Panosso Macedo, Rodrigo and Al Sheikh, Lamis",
    title = "{Pseudospectrum and Black Hole Quasinormal Mode Instability}",
    eprint = "2004.06434",
    archivePrefix = "arXiv",
    primaryClass = "gr-qc",
    doi = "10.1103/PhysRevX.11.031003",
    journal = "Phys. Rev. X",
    volume = "11",
    number = "3",
    pages = "031003",
    year = "2021"
}

@article{Barausse:2014tra,
    author = "Barausse, Enrico and Cardoso, Vitor and Pani, Paolo",
    title = "{Can environmental effects spoil precision gravitational-wave astrophysics?}",
    eprint = "1404.7149",
    archivePrefix = "arXiv",
    primaryClass = "gr-qc",
    doi = "10.1103/PhysRevD.89.104059",
    journal = "Phys. Rev. D",
    volume = "89",
    number = "10",
    pages = "104059",
    year = "2014"
}

@article{Cheung:2021bol,
    author = "Cheung, Mark Ho-Yeuk and Destounis, Kyriakos and Macedo, Rodrigo Panosso and Berti, Emanuele and Cardoso, Vitor",
    title = "{Destabilizing the Fundamental Mode of Black Holes: The Elephant and the Flea}",
    eprint = "2111.05415",
    archivePrefix = "arXiv",
    primaryClass = "gr-qc",
    doi = "10.1103/PhysRevLett.128.111103",
    journal = "Phys. Rev. Lett.",
    volume = "128",
    number = "11",
    pages = "111103",
    year = "2022"
}

@article{Berti:2022xfj,
    author = "Berti, Emanuele and Cardoso, Vitor and Cheung, Mark Ho-Yeuk and Di Filippo, Francesco and Duque, Francisco and Martens, Paul and Mukohyama, Shinji",
    title = "{Stability of the fundamental quasinormal mode in time-domain observations against small perturbations}",
    eprint = "2205.08547",
    archivePrefix = "arXiv",
    primaryClass = "gr-qc",
    doi = "10.1103/PhysRevD.106.084011",
    journal = "Phys. Rev. D",
    volume = "106",
    number = "8",
    pages = "084011",
    year = "2022"
}

@article{Barausse:2014pra,
    author = "Barausse, Enrico and Cardoso, Vitor and Pani, Paolo",
    editor = "Ciani, Giacomo and Conklin, John W. and Mueller, Guido",
    title = "{Environmental Effects for Gravitational-wave Astrophysics}",
    eprint = "1404.7140",
    archivePrefix = "arXiv",
    primaryClass = "astro-ph.CO",
    doi = "10.1088/1742-6596/610/1/012044",
    journal = "J. Phys. Conf. Ser.",
    volume = "610",
    number = "1",
    pages = "012044",
    year = "2015"
}

@article{Abedi:2020ujo,
    author = "Abedi, Jahed and Afshordi, Niayesh and Oshita, Naritaka and Wang, Qingwen",
    title = "{Quantum Black Holes in the Sky}",
    eprint = "2001.09553",
    archivePrefix = "arXiv",
    primaryClass = "gr-qc",
    doi = "10.3390/universe6030043",
    journal = "Universe",
    volume = "6",
    number = "3",
    pages = "43",
    year = "2020"
}

@article{Cardoso:2017cqb,
    author = "Cardoso, Vitor and Pani, Paolo",
    title = "{Tests for the existence of black holes through gravitational wave echoes}",
    eprint = "1709.01525",
    archivePrefix = "arXiv",
    primaryClass = "gr-qc",
    doi = "10.1038/s41550-017-0225-y",
    journal = "Nature Astron.",
    volume = "1",
    number = "9",
    pages = "586--591",
    year = "2017"
}

@article{Oshita:2025ibu,
    author = "Oshita, Naritaka and Berti, Emanuele and Cardoso, Vitor",
    title = "{Unstable Chords and Destructive Resonant Excitation of Black Hole Quasinormal Modes}",
    eprint = "2503.21276",
    archivePrefix = "arXiv",
    primaryClass = "gr-qc",
    reportNumber = "YITP-25-44, RIKEN-iTHEMS-Report-25",
    doi = "10.1103/ht2n-vvvh",
    journal = "Phys. Rev. Lett.",
    volume = "135",
    number = "3",
    pages = "031401",
    year = "2025"
}

@article{Mirbabayi:2018mdm,
    author = "Mirbabayi, Mehrdad",
    title = "{The Quasinormal Modes of Quasinormal Modes}",
    eprint = "1807.04843",
    archivePrefix = "arXiv",
    primaryClass = "gr-qc",
    doi = "10.1088/1475-7516/2020/01/052",
    journal = "JCAP",
    volume = "01",
    pages = "052",
    year = "2020"
}

@article{Kyutoku:2022gbr,
    author = "Kyutoku, Koutarou and Motohashi, Hayato and Tanaka, Takahiro",
    title = "{Quasinormal modes of Schwarzschild black holes on the real axis}",
    eprint = "2206.00671",
    archivePrefix = "arXiv",
    primaryClass = "gr-qc",
    doi = "10.1103/PhysRevD.107.044012",
    journal = "Phys. Rev. D",
    volume = "107",
    number = "4",
    pages = "044012",
    year = "2023"
}

@article{Ianniccari:2024ysv,
    author = "Ianniccari, A. and Iovino, A. J. and Kehagias, A. and Pani, P. and Perna, G. and Perrone, D. and Riotto, A.",
    title = "{Deciphering the Instability of the Black Hole Ringdown Quasinormal Spectrum}",
    eprint = "2407.20144",
    archivePrefix = "arXiv",
    primaryClass = "gr-qc",
    doi = "10.1103/PhysRevLett.133.211401",
    journal = "Phys. Rev. Lett.",
    volume = "133",
    number = "21",
    pages = "211401",
    year = "2024"
}

@article{Okabayashi:2024qbz,
    author = "Okabayashi, Kazumasa and Oshita, Naritaka",
    title = "{Greybody Factors Imprinted on Black Hole Ringdowns. II. Merging Binary Black Holes}",
    eprint = "2403.17487",
    archivePrefix = "arXiv",
    primaryClass = "gr-qc",
    reportNumber = "YITP-24-35, RIKEN-iTHEMS-Report-24",
    month = "3",
    year = "2024"
}

@article{Oshita:2023cjz,
    author = "Oshita, Naritaka",
    title = "{Greybody factors imprinted on black hole ringdowns: An alternative to superposed quasinormal modes}",
    eprint = "2309.05725",
    archivePrefix = "arXiv",
    primaryClass = "gr-qc",
    reportNumber = "YITP-23-111, RIKEN-iTHEMS-Report-23",
    doi = "10.1103/PhysRevD.109.104028",
    journal = "Phys. Rev. D",
    volume = "109",
    number = "10",
    pages = "104028",
    year = "2024"
}

@article{Oshita:2024fzf,
    author = "Oshita, Naritaka and Takahashi, Kazufumi and Mukohyama, Shinji",
    title = "{Stability and instability of the black hole greybody factors and ringdowns against a small-bump correction}",
    eprint = "2406.04525",
    archivePrefix = "arXiv",
    primaryClass = "gr-qc",
    reportNumber = "YITP-24-69, IPMU24-0025, RIKEN-iTHEMS-Report-24",
    doi = "10.1103/PhysRevD.110.084070",
    journal = "Phys. Rev. D",
    volume = "110",
    number = "8",
    pages = "084070",
    year = "2024"
}

@article{Pani:2013pma,
    author = "Pani, Paolo",
    editor = "Cardoso, V. and Gualtieri, L. and Herdeiro, C. and Sperhake, U.",
    title = "{Advanced Methods in Black-Hole Perturbation Theory}",
    eprint = "1305.6759",
    archivePrefix = "arXiv",
    primaryClass = "gr-qc",
    doi = "10.1142/S0217751X13400186",
    journal = "Int. J. Mod. Phys. A",
    volume = "28",
    pages = "1340018",
    year = "2013"
}

@article{Brito:2015oca,
    author = "Brito, Richard and Cardoso, Vitor and Pani, Paolo",
    title = "{Superradiance}: {New Frontiers in Black Hole
    Physics}",
    eprint = "1501.06570",
    archivePrefix = "arXiv",
    primaryClass = "gr-qc",
    doi = "10.1007/978-3-319-19000-6",
    journal = "Lect. Notes Phys.",
    volume = "906",
    pages = "pp.1--237",
    year = "2015"
}

@article{Regge:1957td,
    author = "Regge, Tullio and Wheeler, John A.",
    title = "{Stability of a Schwarzschild singularity}",
    doi = "10.1103/PhysRev.108.1063",
    journal = "Phys. Rev.",
    volume = "108",
    pages = "1063--1069",
    year = "1957"
}

@book{Chandrasekhar:1985kt,
    author = "Chandrasekhar, Subrahmanyan",
    title = "{The mathematical theory of black holes}",
    isbn = "978-0-19-850370-5",
    year = "1985"
}

@article{Berti:2025hly,
    author = "Berti, Emanuele and others",
    title = "{Black hole spectroscopy: from theory to experiment}",
    eprint = "2505.23895",
    archivePrefix = "arXiv",
    primaryClass = "gr-qc",
    month = "5",
    year = "2025"
}

@article{Carullo:2025oms,
    author = "Carullo, Gregorio",
    title = "{Black hole spectroscopy: status report}",
    doi = "10.1007/s10714-025-03408-y",
    journal = "Gen. Rel. Grav.",
    volume = "57",
    number = "5",
    pages = "76",
    year = "2025"
}

@article{LIGOScientific:2016aoc,
    author = "Abbott, B. P. and others",
    collaboration = "LIGO Scientific, Virgo",
    title = "{Observation of Gravitational Waves from a Binary Black Hole Merger}",
    eprint = "1602.03837",
    archivePrefix = "arXiv",
    primaryClass = "gr-qc",
    reportNumber = "LIGO-P150914",
    doi = "10.1103/PhysRevLett.116.061102",
    journal = "Phys. Rev. Lett.",
    volume = "116",
    number = "6",
    pages = "061102",
    year = "2016"
}

@article{LIGOScientific:2025obp,
    collaboration = "LIGO Scientific, VIRGO, KAGRA",
    title = "{Black Hole Spectroscopy and Tests of General Relativity with GW250114}",
    eprint = "2509.08099",
    archivePrefix = "arXiv",
    primaryClass = "gr-qc",
    reportNumber = "LIGO P2500461",
    month = "9",
    year = "2025"
}

@article{LIGOScientific:2025epi,
    author = "Abac, A. G. and others",
    collaboration = "LIGO Scientific, KAGRA, Virgo",
    title = "{GW250114: Testing Hawking{\textquoteright}s Area Law and the Kerr Nature of Black Holes}",
    eprint = "2509.08054",
    archivePrefix = "arXiv",
    primaryClass = "gr-qc",
    reportNumber = "LIGO-P2500421",
    doi = "10.1103/kw5g-d732",
    journal = "Phys. Rev. Lett.",
    volume = "135",
    number = "11",
    pages = "111403",
    year = "2025"
}

@inproceedings{Bambi:2025wjx,
    author = "Bambi, Cosimo and others",
    title = "{Black hole mimickers: from theory to observation}",
    eprint = "2505.09014",
    archivePrefix = "arXiv",
    primaryClass = "gr-qc",
    month = "5",
    year = "2025"
}

@article{Abac:2025saz,
    author = "Abac, Adrian and others",
    title = "{The Science of the Einstein Telescope}",
    eprint = "2503.12263",
    archivePrefix = "arXiv",
    primaryClass = "gr-qc",
    reportNumber = "ET-0036C-25",
    month = "3",
    year = "2025"
}

@article{LIGOScientific:2020iuh,
    author = "Abbott, R. and others",
    collaboration = "LIGO Scientific, Virgo",
    title = "{GW190521: A Binary Black Hole Merger with a Total Mass of $150  M_{\odot}$}",
    eprint = "2009.01075",
    archivePrefix = "arXiv",
    primaryClass = "gr-qc",
    doi = "10.1103/PhysRevLett.125.101102",
    journal = "Phys. Rev. Lett.",
    volume = "125",
    number = "10",
    pages = "101102",
    year = "2020"
}

@article{Dreyer:2003bv,
    author = "Dreyer, Olaf and Kelly, Bernard J. and Krishnan, Badri and Finn, Lee Samuel and Garrison, David and Lopez-Aleman, Ramon",
    title = "{Black hole spectroscopy: Testing general relativity through gravitational wave observations}",
    eprint = "gr-qc/0309007",
    archivePrefix = "arXiv",
    doi = "10.1088/0264-9381/21/4/003",
    journal = "Class. Quant. Grav.",
    volume = "21",
    pages = "787--804",
    year = "2004"
}

@article{Detweiler:1980gk,
    author = "Detweiler, Steven L.",
    title = "{Black holes and gravitational waves. III. the resonant frequencies of rotating holes}",
    doi = "10.1086/158109",
    journal = "Astrophys. J.",
    volume = "239",
    pages = "292--295",
    year = "1980"
}

@article{Berti:2005ys,
    author = "Berti, Emanuele and Cardoso, Vitor and Will, Clifford M.",
    title = "{On gravitational-wave spectroscopy of massive black holes with the space interferometer LISA}",
    eprint = "gr-qc/0512160",
    archivePrefix = "arXiv",
    doi = "10.1103/PhysRevD.73.064030",
    journal = "Phys. Rev. D",
    volume = "73",
    pages = "064030",
    year = "2006"
}

@article{Gossan:2011ha,
    author = "Gossan, S. and Veitch, J. and Sathyaprakash, B. S.",
    title = "{Bayesian model selection for testing the no-hair theorem with black hole ringdowns}",
    eprint = "1111.5819",
    archivePrefix = "arXiv",
    primaryClass = "gr-qc",
    doi = "10.1103/PhysRevD.85.124056",
    journal = "Phys. Rev. D",
    volume = "85",
    pages = "124056",
    year = "2012"
}

@article{Berti:2018vdi,
    author = "Berti, Emanuele and Yagi, Kent and Yang, Huan and Yunes, Nicol\'as",
    title = "{Extreme Gravity Tests with Gravitational Waves from Compact Binary Coalescences: (II) Ringdown}",
    eprint = "1801.03587",
    archivePrefix = "arXiv",
    primaryClass = "gr-qc",
    doi = "10.1007/s10714-018-2372-6",
    journal = "Gen. Rel. Grav.",
    volume = "50",
    number = "5",
    pages = "49",
    year = "2018"
}

@article{LISA:2024hlh,
    author = "Colpi, Monica and others",
    collaboration = "LISA",
    title = "{LISA Definition Study Report}",
    eprint = "2402.07571",
    archivePrefix = "arXiv",
    primaryClass = "astro-ph.CO",
    month = "2",
    year = "2024"
}

@article{Branchesi:2023mws,
    author = "Branchesi, Marica and others",
    title = "{Science with the Einstein Telescope: a comparison of different designs}",
    eprint = "2303.15923",
    archivePrefix = "arXiv",
    primaryClass = "gr-qc",
    reportNumber = "ET-0084A-23",
    doi = "10.1088/1475-7516/2023/07/068",
    journal = "JCAP",
    volume = "07",
    pages = "068",
    year = "2023"
}

@article{ET:2019dnz,
    author = "Maggiore, Michele and others",
    collaboration = "ET",
    title = "{Science Case for the Einstein Telescope}",
    eprint = "1912.02622",
    archivePrefix = "arXiv",
    primaryClass = "astro-ph.CO",
    doi = "10.1088/1475-7516/2020/03/050",
    journal = "JCAP",
    volume = "03",
    pages = "050",
    year = "2020"
}

@article{Kalogera:2021bya,
    author = "Kalogera, Vicky and others",
    title = "{The Next Generation Global Gravitational Wave Observatory: The Science Book}",
    eprint = "2111.06990",
    archivePrefix = "arXiv",
    primaryClass = "gr-qc",
    month = "11",
    year = "2021"
}

@article{Hild:2010id,
    author = "Hild, S. and others",
    title = "{Sensitivity Studies for Third-Generation Gravitational Wave Observatories}",
    eprint = "1012.0908",
    archivePrefix = "arXiv",
    primaryClass = "gr-qc",
    doi = "10.1088/0264-9381/28/9/094013",
    journal = "Class. Quant. Grav.",
    volume = "28",
    pages = "094013",
    year = "2011"
}

@article{LIGOScientific:2016wof,
    author = "Abbott, Benjamin P and others",
    collaboration = "LIGO Scientific",
    title = "{Exploring the Sensitivity of Next Generation Gravitational Wave Detectors}",
    eprint = "1607.08697",
    archivePrefix = "arXiv",
    primaryClass = "astro-ph.IM",
    reportNumber = "LIGO-P1600143",
    doi = "10.1088/1361-6382/aa51f4",
    journal = "Class. Quant. Grav.",
    volume = "34",
    number = "4",
    pages = "044001",
    year = "2017"
}

@article{Berti:2015itd,
    author = "Berti, Emanuele and others",
    title = "{Testing General Relativity with Present and Future Astrophysical Observations}",
    eprint = "1501.07274",
    archivePrefix = "arXiv",
    primaryClass = "gr-qc",
    doi = "10.1088/0264-9381/32/24/243001",
    journal = "Class. Quant. Grav.",
    volume = "32",
    pages = "243001",
    year = "2015"
}

@article{Essick:2017wyl,
    author = "Essick, Reed and Vitale, Salvatore and Evans, Matthew",
    title = "{Frequency-dependent responses in third generation gravitational-wave detectors}",
    eprint = "1708.06843",
    archivePrefix = "arXiv",
    primaryClass = "gr-qc",
    doi = "10.1103/PhysRevD.96.084004",
    journal = "Phys. Rev. D",
    volume = "96",
    number = "8",
    pages = "084004",
    year = "2017"
}

@article{Evans:2023euw,
    author = "Evans, Matthew and others",
    title = "{Cosmic Explorer: A Submission to the NSF MPSAC ngGW Subcommittee}",
    eprint = "2306.13745",
    archivePrefix = "arXiv",
    primaryClass = "astro-ph.IM",
    month = "6",
    year = "2023"
}

@article{Berti:2016lat,
    author = "Berti, Emanuele and Sesana, Alberto and Barausse, Enrico and Cardoso, Vitor and Belczynski, Krzysztof",
    title = "{Spectroscopy of Kerr black holes with Earth- and space-based interferometers}",
    eprint = "1605.09286",
    archivePrefix = "arXiv",
    primaryClass = "gr-qc",
    doi = "10.1103/PhysRevLett.117.101102",
    journal = "Phys. Rev. Lett.",
    volume = "117",
    number = "10",
    pages = "101102",
    year = "2016"
}

@article{Bhagwat:2023jwv,
    author = "Bhagwat, Swetha and Pacilio, Costantino and Pani, Paolo and Mapelli, Michela",
    title = "{Landscape of stellar-mass black-hole spectroscopy with third-generation gravitational-wave detectors}",
    eprint = "2304.02283",
    archivePrefix = "arXiv",
    primaryClass = "gr-qc",
    reportNumber = "ET-0106A-23",
    doi = "10.1103/PhysRevD.108.043019",
    journal = "Phys. Rev. D",
    volume = "108",
    number = "4",
    pages = "043019",
    year = "2023"
}

@article{Bhagwat:2021kwv,
    author = "Bhagwat, Swetha and Pacilio, Costantino and Barausse, Enrico and Pani, Paolo",
    title = "{Landscape of massive black-hole spectroscopy with LISA and the Einstein Telescope}",
    eprint = "2201.00023",
    archivePrefix = "arXiv",
    primaryClass = "gr-qc",
    reportNumber = "ET-0465A-21",
    doi = "10.1103/PhysRevD.105.124063",
    journal = "Phys. Rev. D",
    volume = "105",
    number = "12",
    pages = "124063",
    year = "2022"
}

@article{Vishveshwara:1970zz,
    author = "Vishveshwara, C. V.",
    title = "{Scattering of Gravitational Radiation by a Schwarzschild Black-hole}",
    doi = "10.1038/227936a0",
    journal = "Nature",
    volume = "227",
    pages = "936--938",
    year = "1970"
}

@article{Kokkotas:1999bd,
    author = "Kokkotas, Kostas D. and Schmidt, Bernd G.",
    title = "{Quasinormal modes of stars and black holes}",
    eprint = "gr-qc/9909058",
    archivePrefix = "arXiv",
    doi = "10.12942/lrr-1999-2",
    journal = "Living Rev. Rel.",
    volume = "2",
    pages = "2",
    year = "1999"
}

@article{Berti:2009kk,
    author = "Berti, Emanuele and Cardoso, Vitor and Starinets, Andrei O.",
    title = "{Quasinormal modes of black holes and black branes}",
    eprint = "0905.2975",
    archivePrefix = "arXiv",
    primaryClass = "gr-qc",
    doi = "10.1088/0264-9381/26/16/163001",
    journal = "Class. Quant. Grav.",
    volume = "26",
    pages = "163001",
    year = "2009"
}

@article{Konoplya:2011qq,
    author = "Konoplya, R. A. and Zhidenko, A.",
    title = "{Quasinormal modes of black holes: From astrophysics to string theory}",
    eprint = "1102.4014",
    archivePrefix = "arXiv",
    primaryClass = "gr-qc",
    doi = "10.1103/RevModPhys.83.793",
    journal = "Rev. Mod. Phys.",
    volume = "83",
    pages = "793--836",
    year = "2011"
}

@article{Isi:2019aib,
    author = "Isi, Maximiliano and Giesler, Matthew and Farr, Will M. and Scheel, Mark A. and Teukolsky, Saul A.",
    title = "{Testing the no-hair theorem with GW150914}",
    eprint = "1905.00869",
    archivePrefix = "arXiv",
    primaryClass = "gr-qc",
    reportNumber = "LIGO-P1900135",
    doi = "10.1103/PhysRevLett.123.111102",
    journal = "Phys. Rev. Lett.",
    volume = "123",
    number = "11",
    pages = "111102",
    year = "2019"
}

@article{Franchini:2023eda,
    author = {Franchini, Nicola and V\"olkel, Sebastian H.},
    title = "{Testing General Relativity with Black Hole Quasi-Normal Modes}",
    eprint = "2305.01696",
    archivePrefix = "arXiv",
    primaryClass = "gr-qc",
    month = "5",
    year = "2023"
}

@article{Maggio:2023fwy,
    author = "Maggio, Elisa",
    title = "{Probing the Horizon of Black Holes with Gravitational Waves}",
    eprint = "2310.07368",
    archivePrefix = "arXiv",
    primaryClass = "gr-qc",
    doi = "10.1007/978-3-031-31520-6_9",
    journal = "Lect. Notes Phys.",
    volume = "1017",
    pages = "333--346",
    year = "2023"
}

@article{Cardoso:2021wlq,
    author = "Cardoso, Vitor and Destounis, Kyriakos and Duque, Francisco and Macedo, Rodrigo Panosso and Maselli, Andrea",
    title = "{Black holes in galaxies: Environmental impact on gravitational-wave generation and propagation}",
    eprint = "2109.00005",
    archivePrefix = "arXiv",
    primaryClass = "gr-qc",
    doi = "10.1103/PhysRevD.105.L061501",
    journal = "Phys. Rev. D",
    volume = "105",
    number = "6",
    pages = "L061501",
    year = "2022"
}

@article{Cardoso:2022whc,
    author = "Cardoso, Vitor and Destounis, Kyriakos and Duque, Francisco and Panosso Macedo, Rodrigo and Maselli, Andrea",
    title = "{Gravitational Waves from Extreme-Mass-Ratio Systems in Astrophysical Environments}",
    eprint = "2210.01133",
    archivePrefix = "arXiv",
    primaryClass = "gr-qc",
    doi = "10.1103/PhysRevLett.129.241103",
    journal = "Phys. Rev. Lett.",
    volume = "129",
    number = "24",
    pages = "241103",
    year = "2022"
}

@article{Destounis:2022obl,
    author = "Destounis, Kyriakos and Kulathingal, Arun and Kokkotas, Kostas D. and Papadopoulos, Georgios O.",
    title = "{Gravitational-wave imprints of compact and galactic-scale environments in extreme-mass-ratio binaries}",
    eprint = "2210.09357",
    archivePrefix = "arXiv",
    primaryClass = "gr-qc",
    doi = "10.1103/PhysRevD.107.084027",
    journal = "Phys. Rev. D",
    volume = "107",
    number = "8",
    pages = "084027",
    year = "2023"
}

@article{Biswas:2023ofz,
    author = "Biswas, Shauvik and Singha, Chiranjeeb and Chakraborty, Sumanta",
    title = "{Galactic wormholes: Geometry, stability, and echoes}",
    eprint = "2307.04836",
    archivePrefix = "arXiv",
    primaryClass = "gr-qc",
    doi = "10.1103/PhysRevD.109.064043",
    journal = "Phys. Rev. D",
    volume = "109",
    number = "6",
    pages = "064043",
    year = "2024"
}

@article{Singha:2023lum,
    author = "Singha, Chiranjeeb and Biswas, Shauvik",
    title = "{Galactic pure Lovelock black holes: Geometry, stability, and Hawking temperature}",
    eprint = "2309.01760",
    archivePrefix = "arXiv",
    primaryClass = "gr-qc",
    doi = "10.1103/PhysRevD.109.024043",
    journal = "Phys. Rev. D",
    volume = "109",
    number = "2",
    pages = "024043",
    year = "2024"
}

@article{Gasperin:2021kfv,
    author = "Gasperin, Edgar and Jaramillo, Jos\'e Luis",
    title = "{Energy scales and black hole pseudospectra: the structural role of the scalar product}",
    eprint = "2107.12865",
    archivePrefix = "arXiv",
    primaryClass = "gr-qc",
    doi = "10.1088/1361-6382/ac5054",
    journal = "Class. Quant. Grav.",
    volume = "39",
    number = "11",
    pages = "115010",
    year = "2022"
}

@article{Boyanov:2022ark,
    author = "Boyanov, Valentin and Destounis, Kyriakos and Panosso Macedo, Rodrigo and Cardoso, Vitor and Jaramillo, Jos\'e Luis",
    title = "{Pseudospectrum of horizonless compact objects: A bootstrap instability mechanism}",
    eprint = "2209.12950",
    archivePrefix = "arXiv",
    primaryClass = "gr-qc",
    doi = "10.1103/PhysRevD.107.064012",
    journal = "Phys. Rev. D",
    volume = "107",
    number = "6",
    pages = "064012",
    year = "2023"
}

@article{Sarkar:2023rhp,
    author = "Sarkar, Subhodeep and Rahman, Mostafizur and Chakraborty, Sumanta",
    title = "{Perturbing the perturbed: Stability of quasinormal modes in presence of a positive cosmological constant}",
    eprint = "2304.06829",
    archivePrefix = "arXiv",
    primaryClass = "gr-qc",
    doi = "10.1103/PhysRevD.108.104002",
    journal = "Phys. Rev. D",
    volume = "108",
    number = "10",
    pages = "104002",
    year = "2023"
}

@article{Jaramillo:2022kuv,
    author = "Jaramillo, Jos\'e Luis",
    title = "{Pseudospectrum and binary black hole merger transients}",
    eprint = "2206.08025",
    archivePrefix = "arXiv",
    primaryClass = "gr-qc",
    doi = "10.1088/1361-6382/ac8ddc",
    journal = "Class. Quant. Grav.",
    volume = "39",
    number = "21",
    pages = "217002",
    year = "2022"
}

@article{Arean:2023ejh,
    author = "Are\'an, Daniel and Fari\~na, David Garc\'\i{}a and Landsteiner, Karl",
    title = "{Pseudospectra of holographic quasinormal modes}",
    eprint = "2307.08751",
    archivePrefix = "arXiv",
    primaryClass = "hep-th",
    reportNumber = "IFT-UAM/CSIC-23-091",
    doi = "10.1007/JHEP12(2023)187",
    journal = "JHEP",
    volume = "12",
    pages = "187",
    year = "2023"
}

@article{Cao:2024oud,
    author = "Cao, Li-Ming and Chen, Jia-Ning and Wu, Liang-Bi and Xie, Libo and Zhou, Yu-Sen",
    title = "{The pseudospectrum and spectrum (in)stability of quantum corrected black hole}",
    eprint = "2401.09907",
    archivePrefix = "arXiv",
    primaryClass = "gr-qc",
    reportNumber = "ICTS-USTC/PCFT-24-03",
    month = "1",
    year = "2024"
}

@article{Cownden:2023dam,
    author = "Cownden, Brad and Pantelidou, Christiana and Zilh\~ao, Miguel",
    title = "{The pseudospectra of black holes in AdS}",
    eprint = "2312.08352",
    archivePrefix = "arXiv",
    primaryClass = "gr-qc",
    month = "12",
    year = "2023"
}

@article{Courty:2023rxk,
    author = "Courty, Aubin and Destounis, Kyriakos and Pani, Paolo",
    title = "{Spectral instability of quasinormal modes and strong cosmic censorship}",
    eprint = "2307.11155",
    archivePrefix = "arXiv",
    primaryClass = "gr-qc",
    doi = "10.1103/PhysRevD.108.104027",
    journal = "Phys. Rev. D",
    volume = "108",
    number = "10",
    pages = "104027",
    year = "2023"
}

@inproceedings{Destounis:2023ruj,
    author = "Destounis, Kyriakos and Duque, Francisco",
    title = "{Black-hole spectroscopy: quasinormal modes, ringdown stability and the pseudospectrum}",
    eprint = "2308.16227",
    archivePrefix = "arXiv",
    primaryClass = "gr-qc",
    month = "8",
    year = "2023"
}

@article{Destounis:2023nmb,
    author = "Destounis, Kyriakos and Boyanov, Valentin and Panosso Macedo, Rodrigo",
    title = "{Pseudospectrum of de Sitter black holes}",
    eprint = "2312.11630",
    archivePrefix = "arXiv",
    primaryClass = "gr-qc",
    doi = "10.1103/PhysRevD.109.044023",
    journal = "Phys. Rev. D",
    volume = "109",
    number = "4",
    pages = "044023",
    year = "2024"
}

@article{Cai:2025irl,
    author = "Cai, Rong-Gen and Cao, Li-Ming and Chen, Jia-Ning and Guo, Zong-Kuan and Wu, Liang-Bi and Zhou, Yu-Sen",
    title = "{The pseudospectrum for the Kerr black hole: spin $s=0$ case}",
    eprint = "2501.02522",
    archivePrefix = "arXiv",
    primaryClass = "gr-qc",
    reportNumber = "ICTS-USTC/PCFT-25-02",
    month = "1",
    year = "2025"
}

@article{Gleiser:1995gx,
    author = "Gleiser, Reinaldo J. and Nicasio, Carlos O. and Price, Richard H. and Pullin, Jorge",
    title = "{Second order perturbations of a Schwarzschild black hole}",
    eprint = "gr-qc/9510049",
    archivePrefix = "arXiv",
    reportNumber = "CGPG-95-10-7",
    doi = "10.1088/0264-9381/13/10/001",
    journal = "Class. Quant. Grav.",
    volume = "13",
    pages = "L117--L124",
    year = "1996"
}

@article{Gleiser:1998rw,
    author = "Gleiser, Reinaldo J. and Nicasio, Carlos O. and Price, Richard H. and Pullin, Jorge",
    title = "{Gravitational radiation from Schwarzschild black holes: The Second order perturbation formalism}",
    eprint = "gr-qc/9807077",
    archivePrefix = "arXiv",
    reportNumber = "CGPG-98-7-2",
    doi = "10.1016/S0370-1573(99)00048-4",
    journal = "Phys. Rept.",
    volume = "325",
    pages = "41--81",
    year = "2000"
}

@article{Ioka:2007ak,
    author = "Ioka, Kunihito and Nakano, Hiroyuki",
    title = "{Second and higher-order quasi-normal modes in binary black hole mergers}",
    eprint = "0704.3467",
    archivePrefix = "arXiv",
    primaryClass = "astro-ph",
    doi = "10.1103/PhysRevD.76.061503",
    journal = "Phys. Rev. D",
    volume = "76",
    pages = "061503",
    year = "2007"
}

@article{Nakano:2007cj,
    author = "Nakano, Hiroyuki and Ioka, Kunihito",
    title = "{Second Order Quasi-Normal Mode of the Schwarzschild Black Hole}",
    eprint = "0708.0450",
    archivePrefix = "arXiv",
    primaryClass = "gr-qc",
    doi = "10.1103/PhysRevD.76.084007",
    journal = "Phys. Rev. D",
    volume = "76",
    pages = "084007",
    year = "2007"
}

@article{Brizuela:2009qd,
    author = "Brizuela, David and Martin-Garcia, Jose M. and Tiglio, Manuel",
    title = "{A Complete gauge-invariant formalism for arbitrary second-order perturbations of a Schwarzschild black hole}",
    eprint = "0903.1134",
    archivePrefix = "arXiv",
    primaryClass = "gr-qc",
    doi = "10.1103/PhysRevD.80.024021",
    journal = "Phys. Rev. D",
    volume = "80",
    pages = "024021",
    year = "2009"
}

@article{Pazos:2010xf,
    author = "Pazos, Enrique and Brizuela, David and Martin-Garcia, Jose M. and Tiglio, Manuel",
    title = "{Mode coupling of Schwarzschild perturbations: Ringdown frequencies}",
    eprint = "1009.4665",
    archivePrefix = "arXiv",
    primaryClass = "gr-qc",
    doi = "10.1103/PhysRevD.82.104028",
    journal = "Phys. Rev. D",
    volume = "82",
    pages = "104028",
    year = "2010"
}

@article{Ripley:2020xby,
    author = "Ripley, Justin L. and Loutrel, Nicholas and Giorgi, Elena and Pretorius, Frans",
    title = "{Numerical computation of second order vacuum perturbations of Kerr black holes}",
    eprint = "2010.00162",
    archivePrefix = "arXiv",
    primaryClass = "gr-qc",
    doi = "10.1103/PhysRevD.103.104018",
    journal = "Phys. Rev. D",
    volume = "103",
    pages = "104018",
    year = "2021"
}

@article{Loutrel:2020wbw,
    author = "Loutrel, Nicholas and Ripley, Justin L. and Giorgi, Elena and Pretorius, Frans",
    title = "{Second Order Perturbations of Kerr Black Holes: Reconstruction of the Metric}",
    eprint = "2008.11770",
    archivePrefix = "arXiv",
    primaryClass = "gr-qc",
    doi = "10.1103/PhysRevD.103.104017",
    journal = "Phys. Rev. D",
    volume = "103",
    number = "10",
    pages = "104017",
    year = "2021"
}

@article{Sberna:2021eui,
    author = "Sberna, Laura and Bosch, Pablo and East, William E. and Green, Stephen R. and Lehner, Luis",
    title = "{Nonlinear effects in the black hole ringdown: Absorption-induced mode excitation}",
    eprint = "2112.11168",
    archivePrefix = "arXiv",
    primaryClass = "gr-qc",
    doi = "10.1103/PhysRevD.105.064046",
    journal = "Phys. Rev. D",
    volume = "105",
    number = "6",
    pages = "064046",
    year = "2022"
}

@article{Redondo-Yuste:2023seq,
    author = "Redondo-Yuste, Jaime and Carullo, Gregorio and Ripley, Justin L. and Berti, Emanuele and Cardoso, Vitor",
    title = "{Spin dependence of black hole ringdown nonlinearities}",
    eprint = "2308.14796",
    archivePrefix = "arXiv",
    primaryClass = "gr-qc",
    doi = "10.1103/PhysRevD.109.L101503",
    journal = "Phys. Rev. D",
    volume = "109",
    number = "10",
    pages = "L101503",
    year = "2024"
}

@article{Redondo-Yuste:2023ipg,
    author = "Redondo-Yuste, Jaime and Pere\~niguez, David and Cardoso, Vitor",
    title = "{Ringdown of a dynamical spacetime}",
    eprint = "2312.04633",
    archivePrefix = "arXiv",
    primaryClass = "gr-qc",
    doi = "10.1103/PhysRevD.109.044048",
    journal = "Phys. Rev. D",
    volume = "109",
    number = "4",
    pages = "044048",
    year = "2024"
}

@article{Zhu:2024rej,
    author = "Zhu, Hengrui and others",
    title = "{Nonlinear effects in black hole ringdown from scattering experiments: Spin and initial data dependence of quadratic mode coupling}",
    eprint = "2401.00805",
    archivePrefix = "arXiv",
    primaryClass = "gr-qc",
    doi = "10.1103/PhysRevD.109.104050",
    journal = "Phys. Rev. D",
    volume = "109",
    number = "10",
    pages = "104050",
    year = "2024"
}

@article{Giesler:2019uxc,
    author = "Giesler, Matthew and Isi, Maximiliano and Scheel, Mark A. and Teukolsky, Saul",
    title = "{Black Hole Ringdown: The Importance of Overtones}",
    eprint = "1903.08284",
    archivePrefix = "arXiv",
    primaryClass = "gr-qc",
    doi = "10.1103/PhysRevX.9.041060",
    journal = "Phys. Rev. X",
    volume = "9",
    number = "4",
    pages = "041060",
    year = "2019"
}

@article{Bhagwat:2019dtm,
    author = "Bhagwat, Swetha and Forteza, Xisco Jimenez and Pani, Paolo and Ferrari, Valeria",
    title = "{Ringdown overtones, black hole spectroscopy, and no-hair theorem tests}",
    eprint = "1910.08708",
    archivePrefix = "arXiv",
    primaryClass = "gr-qc",
    doi = "10.1103/PhysRevD.101.044033",
    journal = "Phys. Rev. D",
    volume = "101",
    number = "4",
    pages = "044033",
    year = "2020"
}

@article{DeAmicis:2024eoy,
    author = "De Amicis, Marina and others",
    title = "{Late-time tails in nonlinear evolutions of merging black holes}",
    eprint = "2412.06887",
    archivePrefix = "arXiv",
    primaryClass = "gr-qc",
    month = "12",
    year = "2024"
}

@article{DeAmicis:2024not,
    author = "De Amicis, Marina and Albanesi, Simone and Carullo, Gregorio",
    title = "{Inspiral-inherited ringdown tails}",
    eprint = "2406.17018",
    archivePrefix = "arXiv",
    primaryClass = "gr-qc",
    doi = "10.1103/PhysRevD.110.104005",
    journal = "Phys. Rev. D",
    volume = "110",
    number = "10",
    pages = "104005",
    year = "2024"
}

@article{Baibhav:2023clw,
    author = "Baibhav, Vishal and Cheung, Mark Ho-Yeuk and Berti, Emanuele and Cardoso, Vitor and Carullo, Gregorio and Cotesta, Roberto and Del Pozzo, Walter and Duque, Francisco",
    title = "{Agnostic black hole spectroscopy: Quasinormal mode content of numerical relativity waveforms and limits of validity of linear perturbation theory}",
    eprint = "2302.03050",
    archivePrefix = "arXiv",
    primaryClass = "gr-qc",
    doi = "10.1103/PhysRevD.108.104020",
    journal = "Phys. Rev. D",
    volume = "108",
    number = "10",
    pages = "104020",
    year = "2023"
}

@article{Barack:1998bw,
    author = "Barack, Leor",
    title = "{Late time dynamics of scalar perturbations outside black holes. 2. Schwarzschild geometry}",
    eprint = "gr-qc/9811028",
    archivePrefix = "arXiv",
    doi = "10.1103/PhysRevD.59.044017",
    journal = "Phys. Rev. D",
    volume = "59",
    pages = "044017",
    year = "1999"
}

@article{Price:1972pw,
    author = "Price, Richard H.",
    title = "{Nonspherical Perturbations of Relativistic Gravitational Collapse. II. Integer-Spin, Zero-Rest-Mass Fields}",
    doi = "10.1103/PhysRevD.5.2439",
    journal = "Phys. Rev. D",
    volume = "5",
    pages = "2439--2454",
    year = "1972"
}

@article{Gundlach:1993tp,
    author = "Gundlach, Carsten and Price, Richard H. and Pullin, Jorge",
    title = "{Late time behavior of stellar collapse and explosions: 1. Linearized perturbations}",
    eprint = "gr-qc/9307009",
    archivePrefix = "arXiv",
    reportNumber = "NSF-ITP-93-84",
    doi = "10.1103/PhysRevD.49.883",
    journal = "Phys. Rev. D",
    volume = "49",
    pages = "883--889",
    year = "1994"
}

@article{Abedi:2016hgu,
    author = "Abedi, Jahed and Dykaar, Hannah and Afshordi, Niayesh",
    title = "{Echoes from the Abyss: Tentative evidence for Planck-scale structure at black hole horizons}",
    eprint = "1612.00266",
    archivePrefix = "arXiv",
    primaryClass = "gr-qc",
    doi = "10.1103/PhysRevD.96.082004",
    journal = "Phys. Rev. D",
    volume = "96",
    number = "8",
    pages = "082004",
    year = "2017"
}

@article{Rosato:2025rtr,
    author = "Rosato, Romeo Felice and Pani, Paolo",
    title = "{Universality of late-time ringdown tails}",
    eprint = "2505.08877",
    archivePrefix = "arXiv",
    primaryClass = "gr-qc",
    doi = "10.1103/4yvq-57x6",
    journal = "Phys. Rev. D",
    volume = "112",
    number = "2",
    pages = "024080",
    year = "2025"
}

@article{Leaver:1986gd,
    author = "Leaver, Edward W.",
    title = "{Spectral decomposition of the perturbation response of the Schwarzschild geometry}",
    doi = "10.1103/PhysRevD.34.384",
    journal = "Phys. Rev. D",
    volume = "34",
    pages = "384--408",
    year = "1986"
}

@article{Berti:2006wq,
    author = "Berti, Emanuele and Cardoso, Vitor",
    title = "{Quasinormal ringing of Kerr black holes. I. The Excitation factors}",
    eprint = "gr-qc/0605118",
    archivePrefix = "arXiv",
    doi = "10.1103/PhysRevD.74.104020",
    journal = "Phys. Rev. D",
    volume = "74",
    pages = "104020",
    year = "2006"
}

@article{Silva:2024ffz,
    author = "Silva, Hector O. and Tambalo, Giovanni and Glampedakis, Kostas and Yagi, Kent and Steinhoff, Jan",
    title = "{Quasinormal modes and their excitation beyond general relativity}",
    eprint = "2404.11110",
    archivePrefix = "arXiv",
    primaryClass = "gr-qc",
    doi = "10.1103/PhysRevD.110.024042",
    journal = "Phys. Rev. D",
    volume = "110",
    number = "2",
    pages = "024042",
    year = "2024"
}

@article{Hawking:1975vcx,
    author = "Hawking, S. W.",
    editor = "Gibbons, G. W. and Hawking, S. W.",
    title = "{Particle Creation by Black Holes}",
    doi = "10.1007/BF02345020",
    journal = "Commun. Math. Phys.",
    volume = "43",
    pages = "199--220",
    year = "1975",
    note = "[Erratum: Commun.Math.Phys. 46, 206 (1976)]"
}

@article{LIGOScientific:2021sio,
    author = "Abbott, R. and others",
    collaboration = "LIGO Scientific, VIRGO, KAGRA",
    title = "{Tests of General Relativity with GWTC-3}",
    eprint = "2112.06861",
    archivePrefix = "arXiv",
    primaryClass = "gr-qc",
    reportNumber = "LIGO-P2100275",
    month = "12",
    year = "2021"
}

@article{Destounis:2021lum,
    author = "Destounis, Kyriakos and Macedo, Rodrigo Panosso and Berti, Emanuele and Cardoso, Vitor and Jaramillo, Jos{\'e} Luis",
    title = {{Pseudospectrum of Reissner-Nordstr{\"o}m black holes: Quasinormal mode instability and universality}},
    eprint = "2107.09673",
    archivePrefix = "arXiv",
    primaryClass = "gr-qc",
    doi = "10.1103/PhysRevD.104.084091",
    journal = "Phys. Rev. D",
    volume = "104",
    number = "8",
    pages = "084091",
    year = "2021"
}

@article{Boyanov:2023qqf,
    author = "Boyanov, Valentin and Cardoso, Vitor and Destounis, Kyriakos and Jaramillo, Jos{\'e} Luis and Panosso Macedo, Rodrigo",
    title = "{Structural aspects of the anti{\textendash}de Sitter black hole pseudospectrum}",
    eprint = "2312.11998",
    archivePrefix = "arXiv",
    primaryClass = "gr-qc",
    doi = "10.1103/PhysRevD.109.064068",
    journal = "Phys. Rev. D",
    volume = "109",
    number = "6",
    pages = "064068",
    year = "2024"
}

@article{Cardoso:2024mrw,
    author = "Cardoso, Vitor and Kastha, Shilpa and Panosso Macedo, Rodrigo",
    title = "{Physical significance of the black hole quasinormal mode spectra instability}",
    eprint = "2404.01374",
    archivePrefix = "arXiv",
    primaryClass = "gr-qc",
    doi = "10.1103/PhysRevD.110.024016",
    journal = "Phys. Rev. D",
    volume = "110",
    number = "2",
    pages = "024016",
    year = "2024"
}

@article{Cheung:2022rbm,
    author = "Cheung, Mark Ho-Yeuk and others",
    title = "{Nonlinear Effects in Black Hole Ringdown}",
    eprint = "2208.07374",
    archivePrefix = "arXiv",
    primaryClass = "gr-qc",
    doi = "10.1103/PhysRevLett.130.081401",
    journal = "Phys. Rev. Lett.",
    volume = "130",
    number = "8",
    pages = "081401",
    year = "2023"
}

@article{Mitman:2022qdl,
    author = "Mitman, Keefe and others",
    title = "{Nonlinearities in Black Hole Ringdowns}",
    eprint = "2208.07380",
    archivePrefix = "arXiv",
    primaryClass = "gr-qc",
    doi = "10.1103/PhysRevLett.130.081402",
    journal = "Phys. Rev. Lett.",
    volume = "130",
    number = "8",
    pages = "081402",
    year = "2023"
}

@article{Kehagias:2023ctr,
    author = "Kehagias, Alex and Perrone, Davide and Riotto, Antonio and Riva, Francesco",
    title = "{Explaining nonlinearities in black hole ringdowns from symmetries}",
    eprint = "2301.09345",
    archivePrefix = "arXiv",
    primaryClass = "gr-qc",
    reportNumber = "CERN-TH-2023-076",
    doi = "10.1103/PhysRevD.108.L021501",
    journal = "Phys. Rev. D",
    volume = "108",
    number = "2",
    pages = "L021501",
    year = "2023"
}

@article{Perrone:2023jzq,
    author = "Perrone, Davide and Barreira, Thomas and Kehagias, Alex and Riotto, Antonio",
    title = "{Non-linear black hole ringdowns: An analytical approach}",
    eprint = "2308.15886",
    archivePrefix = "arXiv",
    primaryClass = "gr-qc",
    doi = "10.1016/j.nuclphysb.2023.116432",
    journal = "Nucl. Phys. B",
    volume = "999",
    pages = "116432",
    year = "2024"
}

@article{Cheung:2023vki,
    author = "Cheung, Mark Ho-Yeuk and Berti, Emanuele and Baibhav, Vishal and Cotesta, Roberto",
    title = "{Extracting linear and nonlinear quasinormal modes from black hole merger simulations}",
    eprint = "2310.04489",
    archivePrefix = "arXiv",
    primaryClass = "gr-qc",
    doi = "10.1103/PhysRevD.109.044069",
    journal = "Phys. Rev. D",
    volume = "109",
    number = "4",
    pages = "044069",
    year = "2024",
    note = "[Erratum: Phys.Rev.D 110, 049902 (2024)]"
}

@article{Yi:2024elj,
    author = "Yi, Sophia and Kuntz, Adrien and Barausse, Enrico and Berti, Emanuele and Cheung, Mark Ho-Yeuk and Kritos, Konstantinos and Maselli, Andrea",
    title = "{Nonlinear quasinormal mode detectability with next-generation gravitational wave detectors}",
    eprint = "2403.09767",
    archivePrefix = "arXiv",
    primaryClass = "gr-qc",
    reportNumber = "ET-0081A-24",
    doi = "10.1103/PhysRevD.109.124029",
    journal = "Phys. Rev. D",
    volume = "109",
    number = "12",
    pages = "124029",
    year = "2024"
}

@article{Zhu:2024dyl,
    author = "Zhu, Hengrui and others",
    title = "{Imprints of changing mass and spin on black hole ringdown}",
    eprint = "2404.12424",
    archivePrefix = "arXiv",
    primaryClass = "gr-qc",
    doi = "10.1103/PhysRevD.110.124028",
    journal = "Phys. Rev. D",
    volume = "110",
    number = "12",
    pages = "124028",
    year = "2024"
}

@article{Oshita:2022pkc,
    author = "Oshita, Naritaka",
    title = "{Thermal ringdown of a Kerr black hole: overtone excitation, Fermi-Dirac statistics and greybody factor}",
    eprint = "2208.02923",
    archivePrefix = "arXiv",
    primaryClass = "gr-qc",
    reportNumber = "RIKEN-iTHEMS-Report-22",
    doi = "10.1088/1475-7516/2023/04/013",
    journal = "JCAP",
    volume = "04",
    pages = "013",
    year = "2023"
}

@article{Destounis:2025dck,
    author = "Destounis, Kyriakos and Malato Corr{\^e}a, Mateus and Macedo, Caio F. B. and Panosso Macedo, Rodrigo",
    title = "{Spectral instability of horizonless compact objects within astrophysical environments: The ''exotic'' elephant and the flea}",
    eprint = "2509.16310",
    archivePrefix = "arXiv",
    primaryClass = "gr-qc",
    month = "9",
    year = "2025"
}

@article{Nair:2025anr,
    author = "Nair, Sreejith",
    title = "{Plunge spectra as discriminators of black hole mimickers}",
    eprint = "2509.09986",
    archivePrefix = "arXiv",
    primaryClass = "gr-qc",
    month = "9",
    year = "2025"
}

@article{Siemonsen:2024snb,
    author = "Siemonsen, Nils",
    title = "{Nonlinear Treatment of a Black Hole Mimicker Ringdown}",
    eprint = "2404.14536",
    archivePrefix = "arXiv",
    primaryClass = "gr-qc",
    doi = "10.1103/PhysRevLett.133.031401",
    journal = "Phys. Rev. Lett.",
    volume = "133",
    number = "3",
    pages = "031401",
    year = "2024"
}

@article{Lalanne:2018,
author = {Lalanne, Philippe and Yan, Wei and Vynck, Kevin and Sauvan, Christophe and Hugonin, Jean-Paul},
title = {Light Interaction with Photonic and Plasmonic Resonances},
journal = {Laser \& Photonics Reviews},
volume = {12},
number = {5},
pages = {1700113},
doi = {https://doi.org/10.1002/lpor.201700113},
url = {https://onlinelibrary.wiley.com/doi/abs/10.1002/lpor.201700113},
eprint = {https://onlinelibrary.wiley.com/doi/pdf/10.1002/lpor.201700113},
year = {2018}
}

@phdthesis{Sheikh:2022cud,
    author = "Sheikh, Lamis Al",
    title = "{Scattering resonances and Pseudospectrum : stability and completeness aspects in optical and gravitational systems}",
    reportNumber = "tel-04116011, 2022UBFCK007",
    school = "Institut de Math{\'e}matiques de Bourgogne [Dijon], France",
    year = "2022"
}

@article{Jaramillo:2021tmt,
    author = "Jaramillo, Jos{\'e} Luis and Panosso Macedo, Rodrigo and Sheikh, Lamis Al",
    title = "{Gravitational Wave Signatures of Black Hole Quasinormal Mode Instability}",
    eprint = "2105.03451",
    archivePrefix = "arXiv",
    primaryClass = "gr-qc",
    doi = "10.1103/PhysRevLett.128.211102",
    journal = "Phys. Rev. Lett.",
    volume = "128",
    number = "21",
    pages = "211102",
    year = "2022"
}

@article{Warnick:2024usx,
    author = "Warnick, Claude",
    title = "{(In)stability of de Sitter Quasinormal Mode spectra}",
    eprint = "2407.19850",
    archivePrefix = "arXiv",
    primaryClass = "gr-qc",
    month = "7",
    year = "2024"
}

@misc{Hitrik:2025,
      title={Weyl laws for exponentially small singular values of the $\overline{\partial}$ operator}, 
      author={Michael Hitrik and Johannes Sjöstrand and Martin Vogel},
      year={2025},
      eprint={2505.07292},
      archivePrefix={arXiv},
      primaryClass={math.SP},
      url={https://arxiv.org/abs/2505.07292}, 
}

@article{Zworski:1999,
  title={RESONANCES IN PHYSICS AND GEOMETRY},
  author={Maciej Zworski},
  journal={Notices of the American Mathematical Society},
  year={1999},
  volume={46},
  pages={319-328},
  url={https://api.semanticscholar.org/CorpusID:16822483}
}

@article{Torres:2023nqg,
    author = "Torres, Theo",
    title = "{From Black Hole Spectral Instability to Stable Observables}",
    eprint = "2304.10252",
    archivePrefix = "arXiv",
    primaryClass = "gr-qc",
    doi = "10.1103/PhysRevLett.131.111401",
    journal = "Phys. Rev. Lett.",
    volume = "131",
    number = "11",
    pages = "111401",
    year = "2023"
}
\end{document}